\documentclass[
 reprint,
 amsmath,amssymb,
 aps,
 superscriptaddress,
]{revtex4-2}

\usepackage{graphicx}
\usepackage{bm}

\begin{document}

\preprint{APS/123-QED}

\title{Particle dynamics governed by radiation losses in extreme-field current sheets}

\author{A.~Muraviev}
\email{sashamur@gmail.com}
\affiliation{Institute of Applied Physics of the Russian Academy of Sciences, Nizhny Novgorod 603950, Russia}
\author{A.~Bashinov}
\affiliation{Institute of Applied Physics of the Russian Academy of Sciences, Nizhny Novgorod 603950, Russia}
\author{E.~Efimenko}
\affiliation{Institute of Applied Physics of the Russian Academy of Sciences, Nizhny Novgorod 603950, Russia}
\author{A.~Gonoskov}
\affiliation{Department of Physics, University of Gothenburg, SE-41296 Gothenburg, Sweden}
\affiliation{Lobachevsky State University of Nizhni Novgorod, Nizhny Novgorod 603950, Russia}
\author{I.~Meyerov}
\affiliation{Lobachevsky State University of Nizhni Novgorod, Nizhny Novgorod 603950, Russia}
\author{A.~Sergeev}
\affiliation{Institute of Applied Physics of the Russian Academy of Sciences, Nizhny Novgorod 603950, Russia}

\date{\today}

\begin{abstract}
Particles moving in current sheets under extreme conditions, such as those in the vicinity of pulsars or those predicted on upcoming multipetawatt laser facilities, may be subject to significant radiation losses. We present an analysis of particle motion in fields of a relativistic neutral electron-positron current sheet in the case when radiative effects must be accounted for. In the Landau-Lifshitz radiation reaction force model, when quantum effects are negligible, an analytical solution for particle trajectories is derived. Based on this solution, for the case when quantum effects are significant an averaged quantum solution in the semiclassical approach is obtained. The applicability region of the solutions is determined and analytical trajectories are found to be in good agreement with those of numerical simulations with account for radiative effects. Based on these results we gain new insights into current sheet phenomena expected on upcoming laser facilities.

\end{abstract}


\maketitle

\section{\label{sec:intro}Introduction}
Current sheets are magnetoplasma structures that naturally exist in the Universe. Current sheets with relatively moderate values of the magnetic field strength and particle energies appear, for example, as a result of interaction of solar wind with planetary magnetic fields. Much attention has been paid to these sheets and analytical solutions of equations of particle motion in such structures were obtained \cite{Ness1965,Speiser1965,Speiser1967,Sonnerup1971,Buchner1989,ZelenyiPlPhRep2011,ZelenyiUFN2013,ZelenyiUFN2016}. Characteristic values of magnetic fields and particle energies in this case usually do not require the consideration of radiation losses.

Apart from moderate current sheets there exist extreme ones, for example, in the vicinity of pulsars \cite{Arons2012}. In such sheets the magnetic field and energies of electrons and positrons can be strong enough to ensure abundant hard photon emission and even pair production from photons \cite{Phillipov}. Moreover, thanks to upcoming multipetawatt laser facilities \cite{MPWlasers} extreme sheets may naturally emerge in electron-positron plasma as a result of vacuum breakdown \cite{EfimenkoSR,MuravievJETPL} due to quantum electrodynamic (QED) cascades \cite{BellKirk}. Analysis of current sheet dynamics in this case demands for quantum effects to be taken into account and as a part includes the study of particle dynamics.

Earlier works show that radiative effects can significantly change individual, as well as collective, particle dynamics \cite{ART,EPJ,PukhovPRL,NeitzPRL,Zeld,BulanovPPR,Kirk_2016,GelferSR,Jaroschek2009}. In this paper we investigate theoretically and numerically the influence of radiation losses on dynamics of ultrarelativistic particles in a model extreme current sheet; as a reasonable simplification we assume the magnetic field is fixed (which can often be justified by the relatively high lifetime of current sheets \cite{MuravievJETPL}) and parallel to the current sheet plane. This quasistationary plasma-field configuration is similar to laser excited \cite{EfimenkoSR,MuravievJETPL} and space current sheets \cite{Ness1965, Speiser1965, Runov2003, ZelenyiPlPhRep2011, Arons2012}.

In our study we consider radiation losses of different intensities and consequently within different approaches. In the case when a relativistic particle emits photons frequently and each emitted photon carries away a negligible part of particle energy, it is reasonable to consider radiation losses in the form of the Landau-Lifshitz (LL) force \cite{LL}. In this relatively simple case we derive an approximate analytical solution of equations of motion. In the case when a particle generally loses a large part of its energy in a single act of photon emission, quantum effects significantly affect particle motion and therefore must be taken into account. We modify our solution in order to comply with quantum corrections of power of photon emission \cite{BLP} and obtain an average quantum trajectory of a particle ensemble.

In order to verify our solutions and ranges of their applicability we solve equations of motion numerically with radiation losses within different approaches. The first approach employs the Landau-Lifshitz radiation reaction force. The second one uses the LL force with quantum corrections. The third and the more advanced one is the semiclassical approach \cite{BKF}. This approach assumes probabilistic discrete emissions of photons in accordance with quantum electrodynamics \cite{NikishovRitus,NikishovRitusII} and unperturbed classical Lorentz force-driven motion in between emissions. The semiclassical approach is widely considered as the benchmark (although the terminology may differ) \cite{PoderPRX,ColePRX,Wistisen}. Details of the employed numerical methods for trajectory simulations in the frame of different approaches are given in \cite{BashinovPRE, BashinovPRA}.

The structure of this paper is as follows. In Section~\ref{sec:base}, we establish the setup of the problem, mention previously achieved results for the case without radiation reaction and reformulate them in a form more suitable for our purposes. In Section~\ref{sec:class} we consider radiation reaction in the form of a continuous force of radiative friction in the Landau-Lifshitz form and derive an approximate analytical solution. In Section~\ref{sec:valid} we investigate the region of applicability of this solution and show how it performs outside the theoretical bounds of this region in comparison with a direct numerical solution of equations of motion. In Section~\ref{sec:quant} we provide a method to obtain an average solution in the quantum case based on the solution derived in Section~\ref{sec:class}.

Although current sheets are complicated self-consistent plasma-field structures and particle dynamics should ideally be considered self-consistently with the field generated by all particles, the focus of this paper is the influence of radiation losses on particle dynamics. Therefore, we study the motion of probe particles in a given field modeling a current sheet. Some insights into self-consistent current sheet and plasma dynamics are discussed in Section~\ref{sec:disc}.

\section{\label{sec:base}Base model}

\begin{figure}[b]
\includegraphics[width=\columnwidth]{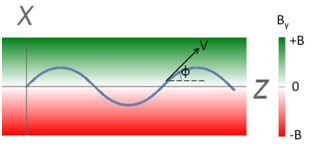}
\caption{\label{fig:TrajField} A sample trajectory of the positron in the $x-z$ plane is in blue. Red-green color shows value of $B_y$.}
\end{figure}

We consider the motion of a positron in a constant inhomogeneous magnetic field with a single non-zero component $B_y(x)$ and an according vector potential $A_z(x)$. Particle motion in a field of such configuration has been studied before in \cite{Speiser1965,Sonnerup1971,ZelenyiUFN2013}, where equations are written in Cartesian coordinates $x$ and $z$. In the present paper we solve equations of motion in coordinates $x$ and $\varphi$, where $\varphi$ is the signed angle between the positron’s velocity and the $z$ axis (see Fig.~\ref{fig:TrajField}), assuming that the positron’s trajectory lies in the $x-z$ plane. We employ such coordinates in order to exploit the analogy with a pendulum oscillating in a gravitational field (see more below). In order to first build a base model, in this section we do not consider radiative effects.

In these coordinates the system of equations can be written as
\begin{equation}
\left\{\begin{array}{l}\dot{\varphi}=-\frac{eB_y(x)}{mc\gamma}=-\frac{e\frac{\partial A_z(x)}{\partial x}}{mc\gamma}
\\
\dot{x}=V(\gamma)\sin{\varphi}\end{array}\right.,\nonumber
\end{equation}
where $e>0$ is the positron charge, $m$ is the positron mass, $c$ is the speed of light and $\gamma$ is the relativistic positron Lorentz-factor. Since the positron’s motion is affected only by the magnetic field, the first integral of motion is the positron’s velocity $V\left(\gamma{}\right)=c\sqrt{1-{\gamma{}}^{-2}}=const$ and therefore $\gamma=const$.

For this system a second integral of motion can be obtained: since the vector-potential $A=(0,0,A_z(x))$ doesn’t depend on $z$, it is evident that $P_z=p_z-eA_z/c=const$.

What is of interest to us here is the properties of particle motion near a null point of the magnetic field. Let us suppose that the magnetic field changes linearly near the null point: $B_y\left(x\right)=kx$, so the vector-potential is a quadratic function: $A_z\left(x\right)=kx^2/2$.

\begin{figure}
\includegraphics[width=\columnwidth]{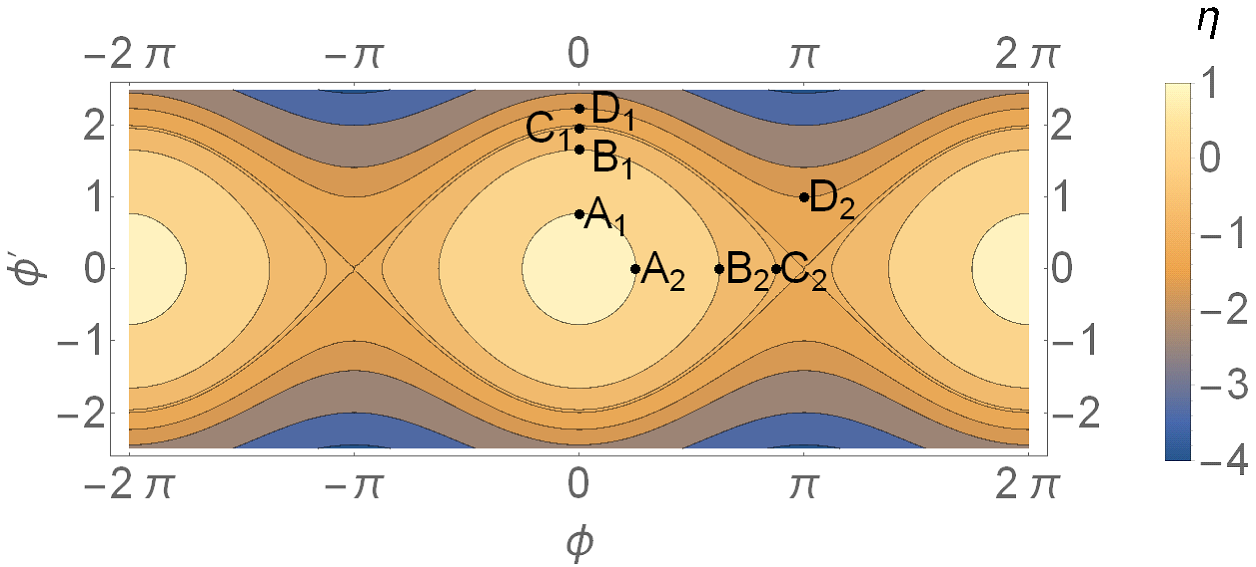}
\caption{\label{fig:PhaseSpace} Phase space given by equations (\ref{systembase}). Certain points are marked, according trajectories are shown in Table~\ref{tab:tablebig}.}
\end{figure}

Then the system of equations can be rewritten as:

\begin{equation}
\label{systembase}
\left\{\begin{array}{l}\dot{\varphi}=-\frac{ek}{mc\gamma}x
\\
\dot{x}=V\sin{\varphi}\end{array}\right.
\end{equation}
or $\ddot{\varphi}+\alpha\left(\gamma{}\right)\sin{\varphi}=0$, where $\alpha(\gamma)=ekV/mc\gamma$ is a constant which depends on the particle's gamma-factor and the slope of the field $k$. We would like to emphasize that in this case the system of equations assumes the exact form of the equations describing an ideal pendulum oscillating in a gravitational field. We use this fact to draw an analogy between positron motion in the field configuration specified above and oscillations of a pendulum. See more in the Appendix. We provide the phase space describing both systems and show the trajectories in real space corresponding to those in the phase space (see Fig.~\ref{fig:PhaseSpace} and Table~\ref{tab:tablebig}).

Dividing $P_z$ by the kinetic momentum $p=mV\gamma$ one can obtain the key dimensionless integral of motion $\eta$, which determines the type of the trajectory (see Table~\ref{tab:tablebig}) of a system with given parameters:
\begin{equation}
\label{eq:eta}
\eta=\cos{\varphi}-\frac{1}{2}\frac{e}{c}\frac{kx^2}{mV\gamma}
\end{equation}

\begin{table*}
\caption{\label{tab:tablebig}Different types of trajectories for a pendulum oscillating in a gravitational field and a positron in a current sheet and the corresponding values of $\eta$. Type A: Low Amplitude Oscillations. Type B: High Amplitude Oscillations. Type C: Near-Separatrix Oscillations. Type D: Circular Motion. $\eta*\approx-0.65$}
\begin{ruledtabular}
\begin{tabular}{cccc}
 & Pendulum & Positron & $\eta$ \\ \hline
A & \includegraphics[width=80 pt]{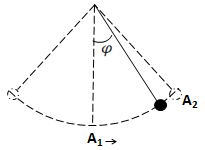} & \includegraphics[width=140 pt]{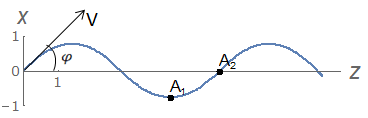} & $0<\eta<1$, $\eta=\cos{\varphi_{max}}$ \\
B & \includegraphics[width=120 pt]{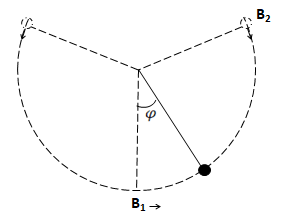} & \includegraphics[width=140 pt]{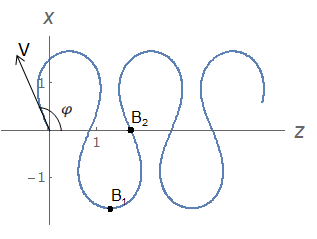} & $\eta*<\eta<0$, $\eta=\cos{\varphi_{max}}$ \\
C & \includegraphics[width=100 pt]{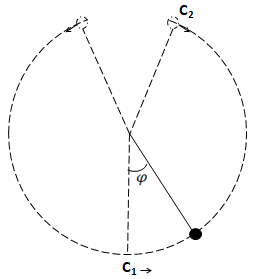} & \includegraphics[width=160 pt]{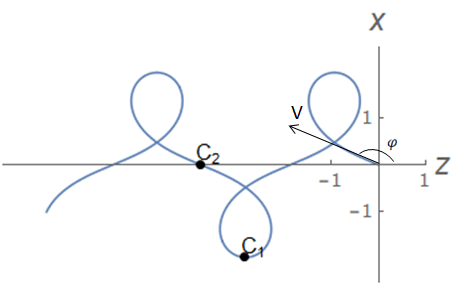} & $-1<\eta<\eta*$, $\eta=\cos{\varphi_{max}}$ \\
D & \includegraphics[width=100 pt]{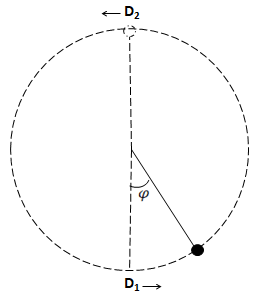} & \includegraphics[width=170 pt]{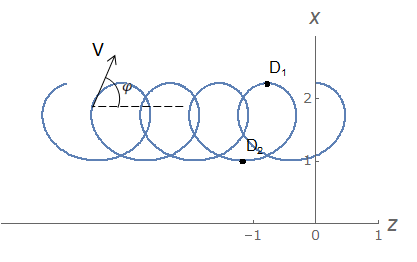} & $\eta<-1$ \\
\end{tabular}
\end{ruledtabular}
\end{table*}

The type and form of trajectory is determined by the value $\eta$. The possible values for $\eta$ (assuming $k>0$) are $-\infty<\eta<1$. Fig.~\ref{fig:PhaseSpace} and Table~\ref{tab:tablebig} show points on the well-known phase space of a pendulum and corresponding points on a trajectory of a positron in a linearly dependent magnetic field for $k>0$.

It can be found by setting $x=0$ and $\cos{\varphi}=-1$ in the equation (\ref{eq:eta}) that $\eta{}{\vert{}}_{sep}=-1$ for a particle on the separatrix. The maximal possible $x$ in the inner region of the separatrix $x_{sep}=2\sqrt{cp/ek}$, which we will call the height of the separatrix, can then be found by setting $\eta{}=\eta\vert_{sep}$ and $\cos{\varphi}=1$.

Values of $\eta$ corresponding to certain trajectories are denoted on Fig.~\ref{fig:PhaseSpace} and Table~\ref{tab:tablebig}. Particularly, it can be seen from (\ref{eq:eta}) that for trajectories that lie inside the separatrix (A-C), and thus cross the $x=0$ line (or the $z$ axis), the exact value of $\eta$ is equal to $\cos{\varphi{}}$ at the instant when the $z$-axis is crossed. Since the angle is maximized on the axis, it can be written that $\eta{}=\cos{\varphi_{max}}$. Accordingly, trajectories type A correspond to values $0<\eta{}<1$, trajectories type B correspond to values $\eta*<\eta{}<0$, trajectories type C: $-1<\eta{}<\eta*$, and trajectories type D: $\eta{}<-1$, where $\eta*\approx-0.65$ corresponds to a closed-curve trajectory shaped similar to the digit 8. The presented classification is similar to the one presented in \cite{Sonnerup1971}.

For trajectories with $\eta$ close to the maximal value $\eta\approx1$ the $\sin{\varphi}$ term in the equations can be linearized similarly to the pendulum equation, and positron motion is close to a harmonic oscillator with frequency $\omega_0=\sqrt{ekV/mc\gamma}$. In the limit $-\eta\gg1$ trajectories resemble a larmor-like gyration with a slow gradient drift in the negative $z$ direction.

To summarize, we rewrote equations for a positron in a given magnetic field in coordinates $(x(t),\varphi(t))$ and we note that in this form the system of equations matches that of an ideal pendulum in a gravitational field. Accordingly, notable analogies were drawn between various entities such as integrals of motion, trajectories, points on trajectories and external parameters.

While these exact trajectories take place in a linearly approximated magnetic field $B_y\sim x$, it is clear that in the general case the $kx^2/2$ term in expression (\ref{eq:eta}) for $\eta$ has to be replaced with the appropriate $A_z(x)$: $\eta=\cos{\varphi}-(e/c)A_z(x)/mV\gamma$. Accordingly, for a certain positron with a known value of $\eta$ the direction of the positron's velocity (determined by $\varphi$) is tied to its coordinate $x$. From this follows that as long as $A_{z}(x)$ is monotonous (meaning there are no additional null points of $B_y$), the above classification of trajectories stands.

\section{\label{sec:class}Radiative recoil: classical approach}

A positron moving along a curvilinear trajectory can emit photons. Based on the preceding work \cite{MuravievJETPL} we allow the magnetic field and particle energy values to be sufficiently high in order for the particles to exhibit radiative recoil. Even though a single impact experienced by the particle as a result of photon emission can be relatively weak, particle motion can be qualitatively modified as a result of a sequence of such acts. While recoil-free motion of particles would be infinite and periodic as described in Sec.1, even recoil insignificant over one period of particle motion due to photon emission may accumulate over multiple periods and have a significant effect on the motion of particles. In the work \cite{MuravievJETPL} current sheets are shown to be formed by ultraintense laser fields. The lifetime of these current sheets was observed to be much larger than the laser wave period, which is in turn much greater than the characteristic times of particle trajectories, so such an accumulation may indeed take place.

In this section we consider particle motion in the field structure with a single non-zero magnetic field component $B_y(x)=kx$ with radiative recoil. In the case when particles emit photons often and they carry away a negligible part of the particle's energy, it is reasonable to consider radiation losses in the form of a continuous Landau-Lifshitz friction force \cite{LL}. The restrictions imposed by this and other assumptions are discussed in Sections (\ref{sec:valid}-\ref{sec:quant}). We also consider only the ultrarelativistic case $p/mc\approx\gamma\gg1$, which allows us to neglect the first and second terms of the LL force \cite{LL}. In our setup this translates to:

\begin{equation}
\label{Frad}
{\vec{F}}_{rad}=-\frac{2e^4\vec{V}}{3m^2c^7}{\gamma}^2V^2k^2x^2
.
\end{equation}

In the system of equations (\ref{systembase}) $\gamma$ was constant as there was no
recoil/friction. Accounting for radiative friction forces us to treat
$\gamma{}$ as another parameter depending on time. Taking into account that
particle energy reduces due to radiative friction given by LL force in Eq.~(\ref{Frad}),
the system of equations (\ref{systembase}) can be rewritten as:

\begin{equation}
\label{systemrad}
\left\{\begin{array}{l}\dot{\varphi}=-\frac{ek}{mc\gamma}x
\\
\dot{x}=V(\gamma)\sin{\varphi} \\
\dot{\gamma}=-\frac{2e^4}{3m^3c^9}{\gamma}^2V^4k^2x^2\end{array}\right.
\end{equation}

We study this system of equations in the case of weak radiative friction, which
allows us to consider the Lorentz-factor of the positron as a slowly changing
parameter. Consequently, the positron's motion at any given moment of time can be
approximated by the solution for the case without radiative friction. Since, as
we know, this motion is periodic, the condition for weakness of radiative friction
can be written as:

\begin{equation}
\label{RRweak}
\frac{\gamma}{\dot{\gamma}}\gg T
,
\end{equation}
where $T$ is the period of motion for the given parameters of the trajectory as described in Section \ref{sec:base}. The condition on the rate of change of trajectory macrocharacteristics is discussed in more detail further in Section \ref{sec:valid}. For further analysis we use solutions without radiative friction as a basis, but we can no longer assume that the prior integral of motion $\eta{}$ remains constant, and therefore must quantify the influence of radiative friction.

We assume $\gamma\gg1$ (meaning $V\approx c$) and employ substitutions $x'=x/c$, $\mu=ek/m\gamma$, $D=2e^5k^3/3m^4c^3$. In this case the system (\ref{systemrad}) can be written as (the prime has been dropped):

\begin{equation}
\label{systemradsubstphi}
\left\{\begin{array}{l}\dot{\varphi}=-\mu x
\\
\dot{x}=\sin{\varphi} \\
\dot{\mu}=Dx^2\end{array}\right.
\end{equation}
Differentiating the second equation of this system yields $\ddot{x}=\dot{\varphi}\cos{\varphi}$. Substituting $\dot{\varphi}$ from the first equation and $\cos{\varphi}$ from (\ref{eq:eta}) yields

\begin{equation}
\label{systemradsubst}
\left\{\begin{array}{l}\ddot{x}=-\mu x(\eta+\frac{\mu x^2}{2})\\
\dot{\mu}=Dx^2\end{array}\right.
,
\end{equation}
where in the new variables
\begin{equation}
\label{eq:etasimp}
\eta=\cos{\varphi}-\mu x^2/2
\end{equation}
.
 $x_{sep}$ can be written as $x_{sep}=2/\sqrt{\mu}$. The term $\mu x^2/2$ can then be written as $2(x/x_{sep})^2$. Also, it is evident that in this case the momentary frequency of oscillations is $\omega=\sqrt{\mu}$.

The solutions of (\ref{systemradsubst}) can be searched for in the form

\begin{equation}
\label{eq:x}
x=Re\left(X(t)e^{i\int\omega(t)dt}\right),
\end{equation}
where $X(t)=x_{max}(t)$ and $\omega\left(t\right)$ are slow real functions. Note that instances when $x(t)=X(t)$ coincide with instances $\varphi(t)=0$, so it can be found from (\ref{eq:etasimp}) that $\mu X^2/2=2(X/x_{sep})^2=1-\eta$.

From the second equation of (\ref{systemradsubst}) $\mu$ can be expressed as $\mu(t)=\mu_0+\int_0^tDx^2dt=\mu_0+(D/2)\int_0^tX^2(t)dt+(D/2)Re\left(\int_0^tX^2(t)e^{2i\int\omega(t)dt}dt\right)$, where the second term represents the time evolution of the "slow" $\left\langle\mu\right\rangle$, and the third term represents the oscillatory part $\tilde{\mu}$. The real part can be expanded, e.g. $Re(Z)=(Z+Z^*)/2$, where * denotes the complex conjugate. After the substitution of $\mu$ and of the assumed form of $x(t)$ (\ref{eq:x}) into the first equation of system (\ref{systemradsubst}), the imaginary part of the term of the resulting equation proportional to $e^{i\int\omega(t)dt}$ can yield: $\dot{\omega}X+2\omega\dot{X}=(DX^3/8\omega)(\eta+3(1-\eta)/8)$. Combining this result with $DX^2/2\approx \dot{\mu}\approx2\omega\dot{\omega}$ gives $d/dt\left(X^\lambda\omega\right)\approx0$ or $X^\lambda\omega\approx const$, where $\lambda=4/(1+5(1-\eta)/8)$. In the case $\varphi\ll1$, equivalent to $\eta\approx1$, $\lambda=4$, so $X^4\omega\approx const$ \footnote{This particular expression can be derived with more ease by assuming $\sin{\varphi}=\varphi$ in (\ref{systemradsubstphi}) and then proceeding with (\ref{eq:x})} can be considered an adiabatic invariant, which in its turn leads to:

\begin{equation}
\label{ThSol}
    \left\{\begin{array}{l}X={X_0\left(1+\frac{t}{\tau}\right)}^{-\frac{1}{10}} \\
    \omega=\omega_0\left(1+\frac{t}{\tau}\right)^{\frac{2}{5}}\end{array}\right.
,
\end{equation}

where $\tau=8{\omega_0}^2/5D{X_0}^2$.

Since $x_{sep}=2/\sqrt{\mu}$ and $\mu={\omega}^2$, $x_{sep}\sim{\left(1+t/\tau\right)}^{-2/5}$. An important consequence is that the decay of the separatrix height $x_{sep}$ is faster than that of the amplitude $X$ of particle oscillations along $x$. Strictly speaking above we derived this only for the case when the particle's oscillations remain close to sinusoidal (\ref{eq:x}), e.g. near the phase space center in the $\varphi\ll1$ region, but it suggests that eventually due to radiative recoil a particle can escape the phase space region enclosed by the separatrix. Let us now show that in fact $X/x_{sep}$ strictly increases without the aforementioned limitation and that the particle can indeed escape.

In the general case it can be shown that $\dot{\eta}\leq0$:\\
$\frac{d}{dt}\left(\cos{\varphi}-\frac{1}{2}\frac{ek}{cp}x^2\right)=\frac{\partial}{\partial\varphi}\left(\cos{\varphi}-\frac{1}{2}\frac{ek}{cp}x^2\right)\dot{\varphi}+\frac{\partial}{\partial x}\left(\cos{\varphi}-\frac{1}{2}\frac{ek}{cp}x^2\right)\dot{x}+\frac{\partial}{\partial p}\left(\cos{\varphi}-\frac{1}{2}\frac{ek}{cp}x^2\right)\dot{p}$. Since $\eta$ is constant in the absence of friction (which means $\dot{p}=0$), the sum of the first two terms is zero, so we obtain: $\frac{d\eta}{dt}=\frac{\partial}{\partial p}\left(\cos{\varphi}-\frac{1}{2}\frac{ek}{cp}x^2\right)\dot{p}=\frac{1}{2}\frac{ek}{cp^2}x^2\left(-\left\vert F_{rad}\right\vert\right)\leq0$.

Now looking at the well-known phase space (Fig.~\ref{fig:PhaseSpace}) and keeping in mind that
$\eta$ decreases the farther the trajectory is from the phase space centre, it is evident that this result in fact means that the ratio $X/x_{sep}$ strictly increases.

\begin{figure}
\includegraphics[width=\columnwidth]{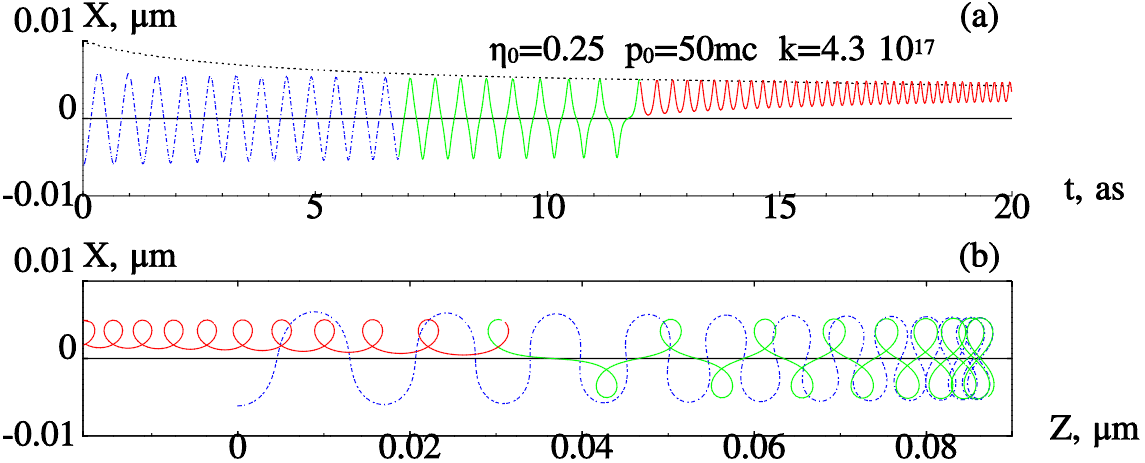}
\caption{\label{fig:Sep} A typical trajectory of a positron experiencing radiative friction in a current sheet. (a) $x(t)$ dependence (b) $x-z$ plane. Colors represent the trajectory transitioning through different trajectory types. Blue (black) dashed – type A-B, green (gray) solid – type C, red (black) solid – type D. Current separatrix height on panel (a) is shown by the dotted black line.}
\end{figure}

We demonstrate this finding in Fig.~\ref{fig:Sep} by numerical modeling of the system of equations (\ref{systemradsubst}) showing a typical particle trajectory along $x(t)$ with its separatrix' height evolution (Fig.~\ref{fig:Sep}a) and the corresponding trajectory on the $x-z$ plane. The qualitative change in motion is clearly seen in Fig.~\ref{fig:Sep}a near the mark $t=12$ as, where the particle stops crossing the $x$ axis, which means that in phase space it has escaped the separatrix.

We note that the type of trajectory and its placing on the phase space can be definitively determined solely by the parameter $\eta$. Remembering that "trapped" trajectories correspond to $-1<\eta<1$ and "escaped"
trajectories correspond to $\eta<-1$ and looking at Fig.~\ref{fig:Sep}, it is evident that this result matches the one obtained in the continuous friction model for $\varphi\ll1$ and $\gamma\gg1$ and confirms that the evolution of particle trajectories can happen only in one direction. In other words, in any model accounting for radiative friction allows particles inside the separatrix to escape to the outer region of the phase space, but not the other way around.

In contrast, "regular" friction (dry and viscous alike) applied to a pendulum results in $\dot{\eta}\geq0$ and all trajectories inevitably settle in one of the equilibrium states (usually the centre).

\section{\label{sec:valid}Theory Validity}

The particle's trajectory in the considered single-component field structure is
described by a system of three first-order differential equations, so its state
is completely defined by three parameters: $\left\{x,\gamma,\varphi\right\}$.
Variable substitution allows us to instead use a different set of variables:
$\{p/mc,\eta,\varphi\}$, where

\begin{equation}
\label{VarSubst}
    \left\{\begin{array}{l}\frac{p}{mc}=\gamma\sqrt{1-{\gamma}^{-2}} \\
    \eta=\frac{P_z}{p}=\cos{\varphi}-\frac{1}{2}\frac{ek}{cp}x^2\end{array}\right.
    .
\end{equation}

In the frictionless case the first two parameters serve as integrals of motion.

It is assumed that radiative friction is weak enough so that the characteristic times of significant change $p/\dot{p}$ and $\mu/\dot{\mu}$ for these parameters are much larger than that of$\ \varphi$. In this case $\varphi$ is considered a quasiperiodic rapidly changing variable, so a
dimension reduction can be performed by eliminating "fast" motion: the "slow" state of the system can be described by just two parameters $p/mc$ and $\eta$.

\begin{figure}[b]
\includegraphics[width=\columnwidth]{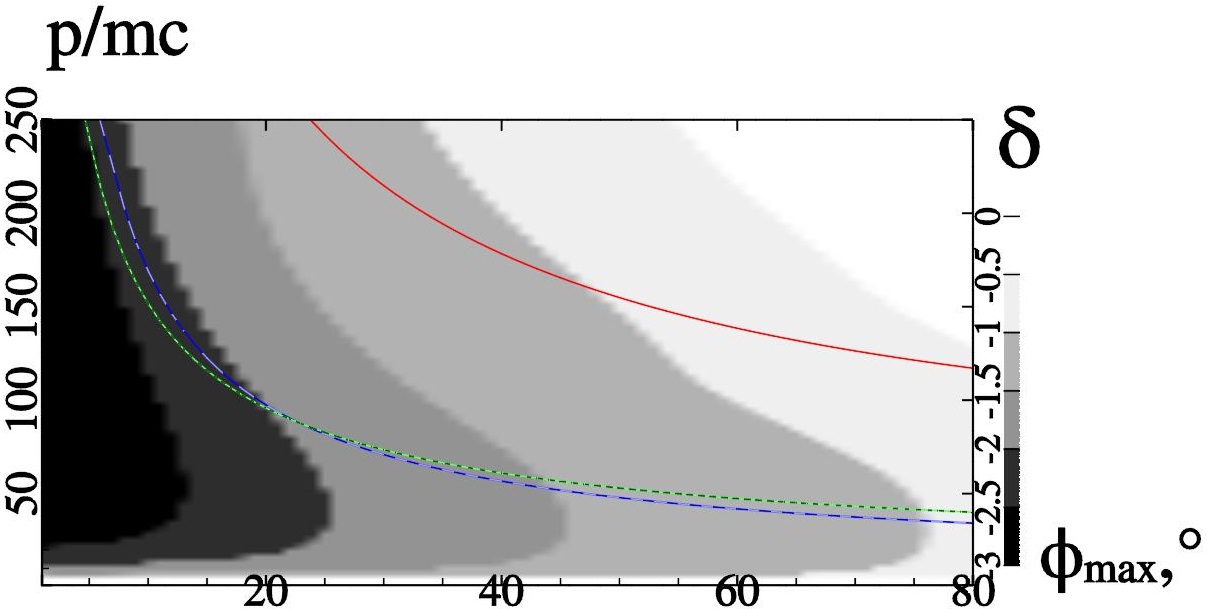}
\caption{\label{fig:DevMap} The discrepancy between theory and numerical solution $\delta$ as a function of $p/mc$ and $\varphi_{max}$, where $\cos{\varphi_{max}}=\eta$. The blue long-dashed line marks the upper edge of the region of weak radiative friction. The green short-dashed line and the solid red line mark the value of the quantum parameter $\chi$ equal to 0.2 and 1, respectively.}
\end{figure}

In order to further study validity and applicability of the developed theory we have compared the rates of change for parameters $p/mc$ and $\eta$, which describe the state of the system, yielded by the theory and by numerical solving for the particle's trajectory using equations (\ref{systemrad}). The numerical simulation is based on the fourth-order Runge-Kutta method with the Landau-Lifshitz force included into equations of motion. In Fig.~\ref{fig:DevMap} we present the map of deviation of these values for the parameter $\eta$\footnote{The map for the parameter $\frac{p}{mc}$ happens to be very similar, so we chose to omit it.}.

Let us denote as ${\dot{\eta}}_{sol}$ the slow (averaged over oscillations of $\varphi$) rate of change of $\eta$ obtained in a numerical solution of the system (\ref{systemrad}), and as ${\dot{\eta}}_{th}$ - the one obtained using the theoretical solution (\ref{ThSol}). Then $\left\vert1-{\dot{\eta}}_{th}/{\dot{\eta}}_{sol}\right\vert$ could be used as a measure of accuracy of the theoretical solution. We use the value $\delta=log_{10}\left\vert1-{\dot{\eta}}_{th}/{\dot{\eta}}_{sol}\right\vert<0$ and plot it as a function of $p/mc$ and $\varphi_{max}$ (where $\cos{\varphi_{max}}=\eta$). The relative difference between the theoretical and numerical result is then equal to ${10}^{\delta}$, which represents the error in the rate of movement of the system on the $(p/mc,\eta)$ parameter plane.

As evident from Fig.~\ref{fig:DevMap}, within the dark area (characteristic values $10\lesssim p/mc\approx\gamma\lesssim150$, $\varphi_{max}\lesssim20^\circ{}$) the derivatives of $\eta_{th}$ and $\eta_{sol}$ differ by less than 1\%. The apparent condition $p/mc>10$ is easily explained by the assumption $\gamma\gg1$ made during derivation of the theory, as is $\varphi_{max}<20^\circ$ by $\varphi\ll1$ rad. The apparently stronger limitation on $\varphi_{max}$ for higher $p/mc$ can be explained by the assumption of weakness of radiative friction (below blue curve in Fig.~\ref{fig:DevMap}). The particular expression derived from (\ref{RRweak}) and (\ref{ThSol}) and used in this Figure is $p/mc=0.1(ek/m)\left(2/D(1-\cos{\varphi_{max}})\right)^{2/5}$, where the value $k\approx4.3\cdot{10}^{17}$ Gs/cm is used, a characteristic value for the problem in \cite{MuravievJETPL}. The red and green curves are discussed in the following section. It should be noted that the position of all three curves is dependent on $k$.

The area marked in black and dark grey can be considered the region of validity of the theoretical results in Section \ref{sec:class}.

\section{\label{sec:quant}Radiative recoil: quantum approach}

In the previous section we showed that the proposed analytical model for particle motion provides results well-matching those obtained by direct solving of the system of differential equations (\ref{systemrad}) (representing the ultrarelativistic case of continuous radiative friction in the LL form) even beyond the theoretical region of applicability. However, in the case of stronger magnetic fields or larger particle energy a particle may lose a significant part of its energy in a single act of photon emission, therefore quantum effects start to affect particle motion. The quantum parameter $\chi=e\hslash/m^3c^4\sqrt{{\left(\varepsilon\vec{E}/c+\vec{p}\times{}\vec{H}\right)}^2-{\left(\vec{p}\cdot\vec{E}\right)}^2}$ is a measure of non-classicality of motion, where $e$ and $m$ are the positron charge and mass, $\hslash$ is the Plank constant, $c$ is the speed of light, $\vec{E}$ and $\vec{H}$ are the electric and magnetic fields and $\varepsilon$ and $\vec{p}$ are the particle's energy and momentum. In our case $\chi\approx\gamma H/H_0$, where $H_0=m^2c^3/e\hslash$ is the Schwinger field. It is usually considered that at $\chi>0.2$ quantum effects start coming into play \cite{BLP,EsirkepovPLA}, however we would like to note that at $\chi=0.2$ the quantum correction already decreases the power of radiation by a factor of two, so the correction may be relevant for even lesser values of $\chi$. We have included the curve
$\chi=0.2$ (in green) in Fig.~\ref{fig:DevMap} to mark the region of applicability of the solution obtained within the LL approach.

When considering the quantum case, first of all, the power of photon emission should be corrected because the LL approach leads to overestimation of radiation losses \cite{BLP}. Second, photon emission has a probabilistic nature, so particles initially in identical conditions may have different trajectories. The more advanced approach applicable in the quantum case is the semiclassical model \cite{BKF}. Within this approach a charged particle is assumed to move classically and friction-free in between instantaneous acts of photon emission. The probability rate and spectrum of photon emission is obtained in quantum electrodynamics \cite{NikishovRitus,NikishovRitusII}. The direction of propagation of the emitted photon is assumed to match that of the parent particle, which is an adequate assumption for the ultrarelativistic case \cite{LL}. Thus the semiclassical method more correctly describes the average power and the stochasticity of photon emission.

Within the semiclassical approach an analytical study in the general case is very complex, in many cases impossible, due to the stochasticity of photon emission. A particle may have a significant probability to lose almost all of its energy in a single act of photon emission, and if a particle may exhibit several regimes of motion, any averaging of parameters may become inaccurate. Moreover, a large and abrupt energy loss can break the condition of slowly varying parameters. These factors become prominent when $\chi\geq1$. For the problem at hand in the current paper this means that a significant portion of particles may abruptly escape out of the separatrix due to a large energy loss and continue motion in a different regime. In Fig.~\ref{fig:DevMap} the curve $\chi=1$ is also presented (in red) and marks the condition for significant stochasticity of photon emission.

However, in the case $\chi\leq1$ averaging of the particle ensemble in the quantum case can be performed analytically. One of the ways is the use of a Fokker-Planck-like equation \cite{NeitzPRL}. Another way is based on averaging of radiative recoil \cite{erber.rmp.1966}. Note that the ratio of the \textit{average} recoil force experienced by the particles in the quantum case to the LL force is a factor (which we will denote as $g$) depending only on $\chi$: $\overline{F_{QC}}/F_C\approx I(\chi)/I_{class}=g(\chi)<1$, where $\overline{F_{QC}}$ and $I(\chi)$ are the averaged force and power given by the semiclassical model and $F_C$ and $I_{class}$ are the classical values \cite{BLP}. Thus a semiclassical model can be reduced to the \textit{corrected} continuous radiation friction model by considering a continuous radiation reaction force equal to the average radiation reaction force given by the semiclassical model: $F_{corr}=\overline{F_{QC}}$.

We implement this correction in the following way. Since we assume that radiative recoil does not change the direction of propagation of the particle, it is evident from the second equation of (\ref{VarSubst}) that $\dot{\eta}\sim\dot{p}$, as well as $\dot{\gamma}\sim\dot{p}=F$ in the ultrarelativistic case. Since the correction has the form of an additional factor $g<1$ in $\dot{p}$ dependant on $\chi$ (and thus, ultimately on parameters $\gamma$ and $\eta$ and oscillation phase $\varphi$), it can be written for instantaneous values that

\begin{eqnarray}
    \dot{\eta}_{corr}(\gamma,\eta,\varphi)=g(\chi(\gamma,\eta,\varphi))\dot{\eta}_{class}(\gamma,\eta,\varphi)
    \nonumber\\
    \dot{\gamma}_{corr}(\gamma,\eta,\varphi)=g(\chi(\gamma,\eta,\varphi))\dot{\gamma}_{class}(\gamma,\eta,\varphi)
\nonumber
\end{eqnarray}

We are interested in the evolution of $\gamma$ and $\eta$ on times much greater than the oscillation period, so the rapid oscillations can be averaged out: we will denote with an $\overline{overline}$ values averaged over a period of rapid oscillations of $\varphi$. In this way, the averaged value is a function of only parameters $\gamma$ and $\eta$.

$\chi(t)$ can be viewed as a product of the slowly changing envelope $\chi_{max}(\gamma(t),\eta(t))$ and fast oscillations $f(\varphi(t))$. In this expression $f(\varphi(t))$ is a quasiperiodic function, so $g(\chi(\gamma,\eta,\varphi))$ is also
quasiperiodic.

In this way, values yielded by the \textit{corrected} model can be computed as
\begin{eqnarray}
\eta_{corr}(t)\approx\eta(t=0)+\int_0^t\overline{g(\chi){\frac{d\eta}{dt}}_{class}}dt
\nonumber\\
\gamma_{corr}(t)\approx\gamma(t=0)+\int_0^t\overline{g(\chi){\frac{d\gamma}{dt}}_{class}}dt
\end{eqnarray}
In simpler words these curves can be obtained by stretching every $dt$ in the
classical curves by a factor of $1/g(\chi(t))$.

\subsection{\label{sec:AvQSol}Averaged Quantum Solution}

We have performed a series of modeling of particle dynamics using the semiclassical approach with the help of our code, based on the Runge-Kutta method, to solve equations of motion and the Monte-Carlo method to simulate random acts of photon emission \cite{BashinovPRE}. In order to be able to compare semiclassical results with continuous-friction results, for each of the initial conditions considered (an initial condition is defined by $\left\vert p\right\vert$ and $\eta$) 100 semiclassical trajectories were analyzed and the average parameters $\left\vert p\right\vert$ and $\eta$ were computed for each moment of time $t$. These results were compared with the results obtained in the classical LL model of continuous radiative friction and the \textit{corrected} continuous radiative friction model.

\begin{figure}
\includegraphics[width=\columnwidth]{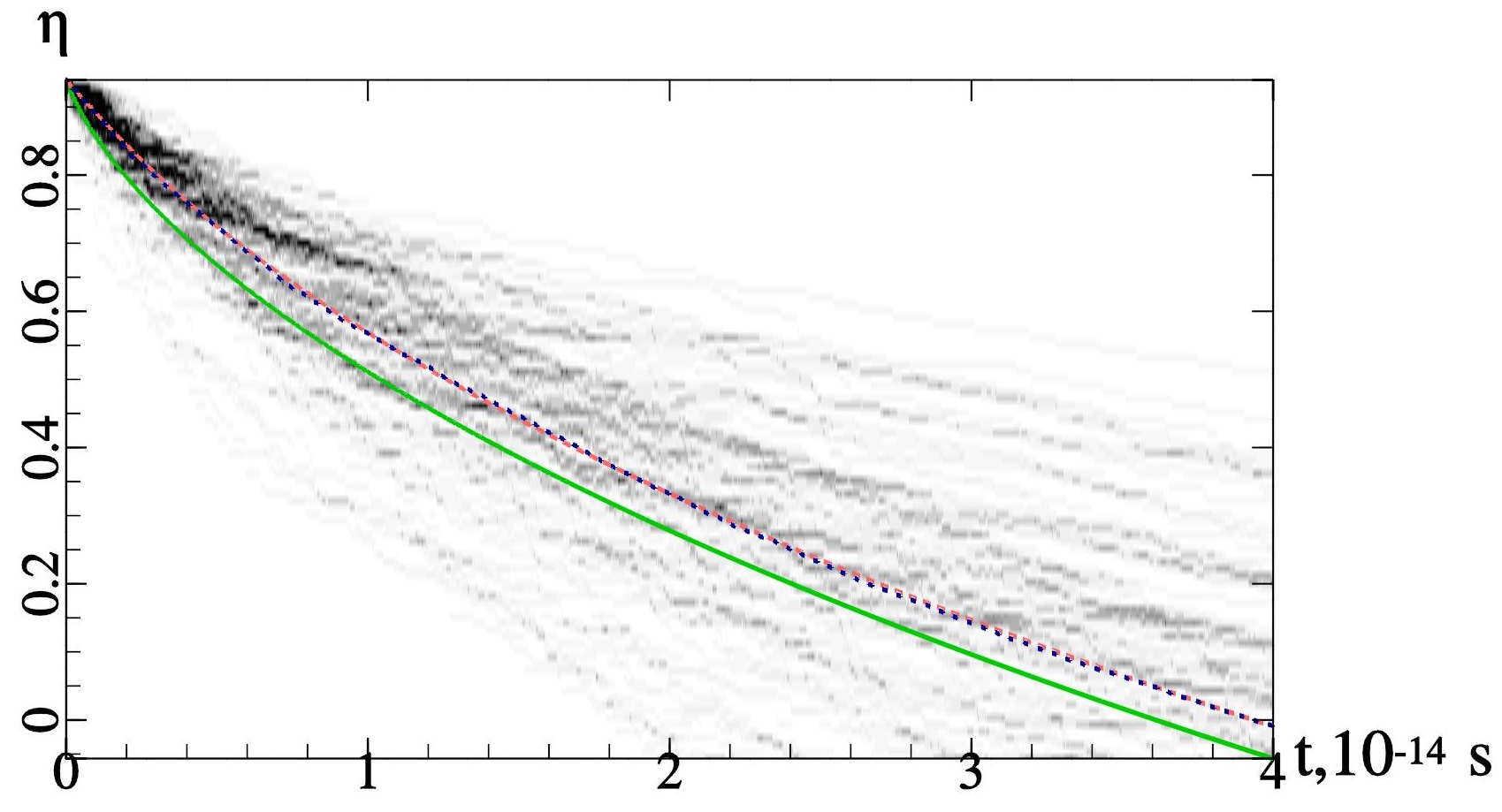}
\caption{\label{fig:Eta_t} An example $\eta(t)$ using different models. Classical LL model - green solid line, Corrected LL model - red long-dashed line, Semiclassical case (averaged) – blue dotted line, Semiclassical case (100 particles) – black. The initial parameters used are $p/mc\approx176$, $\varphi_{max}=20^\circ$}
\end{figure}

An example of evolution of $\eta$ obtained using different models is presented in Fig.~\ref{fig:Eta_t}. As predicted, a clearly observable difference is found in the evolution of slow parameters of particles (in this case $\eta$) between the averaged semiclassical case (blue) and the classical LL friction (green). However, it was also shown that this difference is substantially negated by using the \textit{corrected}  LL friction model (red).

Particularly, it was obtained that for values of $\chi\sim1$ parameters of trajectories ($p$ and $\eta$) averaged over a large number of random realizations are well approximated by the \textit{corrected} LL friction model. This result is substantial because although individual trajectories at such values of $\chi$ feature highly non-classical individual dynamics, the average dynamics allows for analytical description. As a quantitative measure of the effectiveness of such a correction we offer the average value of $\left\vert\eta_{av}-\eta_{corr}\right\vert/\left\vert\eta_{av}-\eta_{LL}\right\vert$, which in the case of quite strong stochasticity $\chi_{init}=0.5$ amounts to $\lesssim 0.1$.

As such, the described above method can be used to construct an averaged quantum solution from the classical solution obtained with the LL model.

\section{\label{sec:disc}Discussion}

In this section we try to apply our new knowledge on particle trajectories to more realistic cases of QED-generated current sheets by ultraintense laser fields \cite{MuravievJETPL}. The problem covered in that paper based on self-consistent electron-positron plasma dynamics differs substantially from the single particle approach developed in the current paper as it includes electron-positron pair generation and self-consistent generation of magnetic and electric fields by the motion of  particles constituting the current sheet. Here we provide some insights from the results gained in this research that can provide better understanding of processes during self-consistent current sheet formation. This problem obviously requires thorough analysis so here we provide simple qualitative considerations that can nevertheless be useful.

First, we would like to consider the evolution of the $z$-directed current given by the particle's motion. The dependence of the current given by a single particle on $\eta$ without radiative friction has been studied in \cite{Sonnerup1971}. In the current paper it was shown that radiative friction causes trajectories of particles to decrease in amplitude, for particles with $\gamma\gg1$ and near the phase space center ($\varphi\ll1$) this was shown both analytically and numerically. This means that the width of the current profile $j_z(x)$ of such particles also decreases with time. Interestingly, for $\varphi\ll1$ this can happen without significant loss of total current: indeed, the current $I_z$ is proportional to $V\overline{\cos{\varphi}}$. Despite the loss of energy due to radiative recoil, $\frac{dV}{dt}$ can be made small by considering high values of $\gamma$ so that $V\approx c$. $\varphi\ll1$ ($\cos{\varphi}\approx1$) makes it possible to have negligible $\frac{d\overline{\cos{\varphi}}}{dt}$, and consequently, $\frac{dI_z}{dt}$, while the current profile width (proportional to $\sqrt{\gamma(1-\cos{\varphi_{max}})}$) can decrease significantly.

Second, we would like to briefly discuss the effects of electron-positron pair generation bringing new particles into the fold on the process of formation of current sheets. As follows from Section \ref{sec:class}, each particle in the presented field configuration has an indefinitely restricted (along the $x$ axis) region in regular space outside of the bounds set by its current amplitude of oscillations. This amplitude can only decrease and therefore the particle never escapes this region. The particle trajectory envelope has a shape of a narrowing cone, and this property is not broken even when the particle crosses the separatrix, see Fig.~\ref{fig:Sep}.
Consequently, during generation of new particles that occurs in such a system neither parent nor the offspring particles can escape further from the current sheet than designed by their respective initial trajectories (unperturbed by radiative friction). This may lead to accumulation of particles near the current sheet and the consequent increase in particle density $\overline{n(x)}$ (averaged over a period of time larger than the characteristic period of fast motion).

Finally, current sheets manifesting in astrophysical circumstances often exhibit more complex field structures, including other components of magnetic and electric fields. A commonly studied configuration includes a component $E_z$ of the electric field along the sheet in the direction of propagation of positrons as described above. Such a field configuration is of particular interest to us since it resembles the one forming in work \cite{MuravievJETPL}. For simplicity we assume that the additional electric field $E_z(x,t)>0$ is constant in time and homogeneous in space. The homogeneity assumption can be justified by $\frac{\partial E_z}{\partial x}/E_z\ll1/d$ (where $d$ is the width of the sheet) at the start of the process, which is usually the case.

Let us discuss how such an electric field would affect particle trajectories. Particles that would otherwise be of trajectory type D, instead of slowly
drifting along $z$ in the negative direction with minimal average current due to the gradient of the magnetic field $B_y$, would instead drift in crossed fields $E_z$ and $B_y$ towards the $x=0$ plane. Particles that had already been swept towards or those initially close to this plane (other trajectory types, especially type A) would engage in similar motion as shown in Section \ref{sec:base}, except they are further accelerated towards the $+z$ direction.

Furthermore, a strong electric field yields additional pair production. These newly born particles are too swept towards the $x=0$ plane in the crossed fields and then accelerated towards $z$, which creates an additional $z$-directed current with the characteristic width defined and limited by the separatrix height $x_{sep}$ depending on the particles' characteristic gamma (see Section \ref{sec:base}). This current further increases the slope $\frac{\partial B_y}{\partial x}$, which, in turn, further narrows $x_{sep}$, creating a positive feedback, so this problem can no longer be seen as a problem in fixed fields.

While our model does not allow to search for a limit to this positive feedback process, the natural limitation would be the depletion of the electric field $E_z$ serving as a source of energy for the particles. As observed in \cite{MuravievJETPL}, the electric field in the current sheet vicinity is indeed eventually absorbed or relaxed and only noise values are present, at which point the regime of slow self-consistent evolution of currents and the magnetic field begins.

\section{\label{sec:conc}Conclusion}

Motion of ultrarelativistic charged particles in neutral current sheets taking into account radiation reaction was considered. Their phase space was studied and analytical solutions were obtained in the approximation near the phase space center. It was demonstrated both analytically and numerically that a key parameter (serving in the frictionless case as an integral of motion) $\eta$ strictly decreases as the result of radiative firction. Since this parameter solely defines the type of a particle’s trajectory, this defines the path of evolution of particles’ trajectory types: from current carrying trajectories in the $+z$ direction along the sheet to Larmor-like gyration with a weak drift in the $-z$ direction. Analytical solutions were compared against numerical solutions of the system of differential equations featuring radiative friction in the classical Landau-Lifshitz form and found to be a match within $1\%$ inside the theoretical region of applicability of the analytics and within $10\%$ well outside of it. A comparison of models featuring continuous radiative friction against semiclassical models was performed, it was shown that the usage of the corrected LL model significantly reduces the error of the continuous models versus the results of the semiclassical model averaged over a large number of realizations. As a result, an analytical description of averaged parameters of particle trajectories is possible in the semiclassical case. Finally, the influence of radiative friction on individual particles’ motion was discussed in scope of self-consistent current sheets formation by ultraintense laser fields considered previously in \cite{MuravievJETPL}.

\begin{acknowledgments}
The research was supported by the Ministry of Science and Higher Education of the Russian Federation, state assignment for the Institute of Applied Physics RAS, project No. 0030-2021-0012.
A.M. acknowledges that the work was supported by the Foundation for the Advancement of Theoretical Physics and Mathematics "BASIS".
A.G. and I.M. acknowledge the support of the Lobachevsky State University of Nizhny Novgorod academic excellence program (Project 5-100).
The authors acknowledge the use of computational resources provided by the Joint Supercomputer Center of the Russian Academy of Sciences and by the Swedish National Infrastructure for Computing (SNIC).
\end{acknowledgments}

\appendix*
\section{\label{Appendix}Analogy between positron in current sheet and pendulum}

The system of equations (\ref{systembase}) assumes the exact form of the equations describing an ideal pendulum oscillating in a gravitational field, which is assumed to be a mass $m$ (we intentionally use the same designation as the positron's mass) suspended on a massless rod of length $l$ under influence of gravitational acceleration $g$. We use this fact to draw an analogy between positron motion in the field configuration specified above and oscillations of a pendulum. No kind of energy loss or friction is considered in this section. The phase space describing both systems and the corresponding trajectories in real space can be found in (see Fig.~\ref{fig:PhaseSpace} and Table~\ref{tab:tablebig}).

The corresponding physical values and equations for these two problems can be seen in Table~\ref{tab:tablesmall}. First of all, the external conditions driving the system are determined by the values $g$ for the pendulum and $k=\frac{dB_y}{dx}$ for the positron. Second, the length of the pendulum $l$ and speed $V$ or kinetic energy $E_k$ of the positron are both crucial properties of the system that determine its dynamics and are both constant throughout its evolution. $E_k$ or $V$ or $\gamma$ (any one of the three values can be expressed through any one of the others) can be considered as the first integral of motion for the positron. Third, the $z$-component of generalized momentum $\vec{P}$ - the second integral of motion for the positron - corresponds to the full mechanical energy $E_M$ of the pendulum. Dividing $P_z$ by the kinetic momentum $p=mV\gamma$ and $-E_M$ by the maximal potential energy of pendulum (see Table~\ref{tab:tablesmall}), one can obtain the key dimensionless integral of motion $\eta$ (see last line of Table \ref{tab:tablesmall}), which determines the type of the trajectory (see Table~\ref{tab:tablebig}) of a system with given parameters. Note that in the case of the pendulum a higher velocity $V$ (or kinetic energy $E_K$) of the pendulum results in a lower $\eta{}$: $\frac{\partial\eta}{\partial E_K}\leq0$, while for the positron is it the opposite: $\frac{\partial\eta}{\partial E_K}\geq0$.

Beside the constant in time external conditions and integrals of motion the analogy also extends to time-dependant variables. The angle $\varphi$ formed between the pendulum and the vertical axis fully corresponds to the angle $\varphi$ between the positron's velocity and the axis $z$, so in this case we intentionally use the same designation for these angles. The $x$ coordinate for the positron does not have a direct analogy, but it is proportional to $\dot{\varphi}$ of both the positron and the pendulum.

\begin{table}[h]
\caption{\label{tab:tablesmall}
Corresponding physical values and equations for a pendulum oscillating in a gravitational field and a positron in a current sheet}
\begin{ruledtabular}
\begin{tabular}{cccccccc}
 &Pendulum &Positron in\\
 & & current sheet\\
\hline
Ext. Parameter & $g$ & $k$ \\
Int. Property & $l=const$ & $E_k=const$ \\
Cyclic Variable & $\varphi$ & $\varphi$ \\ 
Ang. Velocity & $\dot{\varphi}$ & $\dot{\varphi}$ \\
Frequency\footnote{of infinitesimal osillations} & $\omega_0=\sqrt{\frac{g}{l}}$ & $\omega_0=\sqrt{\frac{ekV}{mc\gamma}}$\\
Diff. Equation & $\ddot{\varphi}+\frac{g}{l}\sin{\varphi}=0$ & $\ddot{\varphi}+\frac{ekV}{mc\gamma}\sin{\varphi}=0$ \\
{$E_M$} / {$P_z$} &  $\frac{1}{2}ml^2{\dot{\varphi}}^2-mgl\cos{\varphi}$ & $mV\gamma{}\cos{\varphi-\frac{1}{2}\frac{e}{c}kx^2}$ \\
$\eta$ & $\cos{\varphi}-\frac{1}{2}\frac{V^2}{gl}$ & $\cos{\varphi}-\frac{1}{2}\frac{e}{c}\frac{kx^2}{mV\gamma}$

\end{tabular}
\end{ruledtabular}
\end{table}

\bibliography{bibliography}

\providecommand{\noopsort}[1]{}\providecommand{\singleletter}[1]{#1}%
\begin{thebibliography}{38}%
\makeatletter
\providecommand \@ifxundefined [1]{%
 \@ifx{#1\undefined}
}%
\providecommand \@ifnum [1]{%
 \ifnum #1\expandafter \@firstoftwo
 \else \expandafter \@secondoftwo
 \fi
}%
\providecommand \@ifx [1]{%
 \ifx #1\expandafter \@firstoftwo
 \else \expandafter \@secondoftwo
 \fi
}%
\providecommand \natexlab [1]{#1}%
\providecommand \enquote  [1]{``#1''}%
\providecommand \bibnamefont  [1]{#1}%
\providecommand \bibfnamefont [1]{#1}%
\providecommand \citenamefont [1]{#1}%
\providecommand \href@noop [0]{\@secondoftwo}%
\providecommand \href [0]{\begingroup \@sanitize@url \@href}%
\providecommand \@href[1]{\@@startlink{#1}\@@href}%
\providecommand \@@href[1]{\endgroup#1\@@endlink}%
\providecommand \@sanitize@url [0]{\catcode `\\12\catcode `\$12\catcode
  `\&12\catcode `\#12\catcode `\^12\catcode `\_12\catcode `\%12\relax}%
\providecommand \@@startlink[1]{}%
\providecommand \@@endlink[0]{}%
\providecommand \url  [0]{\begingroup\@sanitize@url \@url }%
\providecommand \@url [1]{\endgroup\@href {#1}{\urlprefix }}%
\providecommand \urlprefix  [0]{URL }%
\providecommand \Eprint [0]{\href }%
\providecommand \doibase [0]{http://dx.doi.org/}%
\providecommand \selectlanguage [0]{\@gobble}%
\providecommand \bibinfo  [0]{\@secondoftwo}%
\providecommand \bibfield  [0]{\@secondoftwo}%
\providecommand \translation [1]{[#1]}%
\providecommand \BibitemOpen [0]{}%
\providecommand \bibitemStop [0]{}%
\providecommand \bibitemNoStop [0]{.\EOS\space}%
\providecommand \EOS [0]{\spacefactor3000\relax}%
\providecommand \BibitemShut  [1]{\csname bibitem#1\endcsname}%
\let\auto@bib@innerbib\@empty
%</preamble>
\bibitem [{\citenamefont {Ness}(1965)}]{Ness1965}%
  \BibitemOpen
  \bibfield  {author} {\bibinfo {author} {\bibfnamefont {N.~F.}\ \bibnamefont
  {Ness}},\ }\href {\doibase https://doi.org/10.1029/JZ070i013p02989}
  {\bibfield  {journal} {\bibinfo  {journal} {Journal of Geophysical Research}\
  }\textbf {\bibinfo {volume} {70}},\ \bibinfo {pages} {2989} (\bibinfo {year}
  {1965})}\BibitemShut {NoStop}%
\bibitem [{\citenamefont {Speiser}(1965)}]{Speiser1965}%
  \BibitemOpen
  \bibfield  {author} {\bibinfo {author} {\bibfnamefont {T.~W.}\ \bibnamefont
  {Speiser}},\ }\href {\doibase https://doi.org/10.1029/JZ070i017p04219}
  {\bibfield  {journal} {\bibinfo  {journal} {Journal of Geophysical Research
  (1896-1977)}\ }\textbf {\bibinfo {volume} {70}},\ \bibinfo {pages} {4219}
  (\bibinfo {year} {1965})}\BibitemShut {NoStop}%
\bibitem [{\citenamefont {Speiser}(1967)}]{Speiser1967}%
  \BibitemOpen
  \bibfield  {author} {\bibinfo {author} {\bibfnamefont {T.~W.}\ \bibnamefont
  {Speiser}},\ }\href {\doibase https://doi.org/10.1029/JZ072i015p03919}
  {\bibfield  {journal} {\bibinfo  {journal} {Journal of Geophysical Research
  (1896-1977)}\ }\textbf {\bibinfo {volume} {72}},\ \bibinfo {pages} {3919}
  (\bibinfo {year} {1967})}\BibitemShut {NoStop}%
\bibitem [{\citenamefont {Sonnerup}(1971)}]{Sonnerup1971}%
  \BibitemOpen
  \bibfield  {author} {\bibinfo {author} {\bibfnamefont {B.~U.~O.}\
  \bibnamefont {Sonnerup}},\ }\href {\doibase
  https://doi.org/10.1029/JA076i034p08211} {\bibfield  {journal} {\bibinfo
  {journal} {Journal of Geophysical Research (1896-1977)}\ }\textbf {\bibinfo
  {volume} {76}},\ \bibinfo {pages} {8211} (\bibinfo {year}
  {1971})}\BibitemShut {NoStop}%
\bibitem [{\citenamefont {Büchner}\ and\ \citenamefont
  {Zelenyi}(1989)}]{Buchner1989}%
  \BibitemOpen
  \bibfield  {author} {\bibinfo {author} {\bibfnamefont {J.}~\bibnamefont
  {Büchner}}\ and\ \bibinfo {author} {\bibfnamefont {L.~M.}\ \bibnamefont
  {Zelenyi}},\ }\href {\doibase https://doi.org/10.1029/JA094iA09p11821}
  {\bibfield  {journal} {\bibinfo  {journal} {Journal of Geophysical Research:
  Space Physics}\ }\textbf {\bibinfo {volume} {94}},\ \bibinfo {pages} {11821}
  (\bibinfo {year} {1989})}\BibitemShut {NoStop}%
\bibitem [{\citenamefont {Zelenyi}\ \emph {et~al.}(2011)\citenamefont
  {Zelenyi}, \citenamefont {Malova}, \citenamefont {Artemyev}, \citenamefont
  {Popov},\ and\ \citenamefont {Petrukovich}}]{ZelenyiPlPhRep2011}%
  \BibitemOpen
  \bibfield  {author} {\bibinfo {author} {\bibfnamefont {L.~M.}\ \bibnamefont
  {Zelenyi}}, \bibinfo {author} {\bibfnamefont {H.~V.}\ \bibnamefont {Malova}},
  \bibinfo {author} {\bibfnamefont {A.~V.}\ \bibnamefont {Artemyev}}, \bibinfo
  {author} {\bibfnamefont {V.~Y.}\ \bibnamefont {Popov}}, \ and\ \bibinfo
  {author} {\bibfnamefont {A.~A.}\ \bibnamefont {Petrukovich}},\ }\href
  {\doibase https://doi.org/10.1134/S1063780X1102005X} {\bibfield  {journal}
  {\bibinfo  {journal} {Plasma Phys. Rep.}\ }\textbf {\bibinfo {volume} {37}},\
  \bibinfo {pages} {118–160} (\bibinfo {year} {2011})}\BibitemShut {NoStop}%
\bibitem [{\citenamefont {Zelenyi}\ \emph {et~al.}(2013)\citenamefont
  {Zelenyi}, \citenamefont {Neishtadt}, \citenamefont {Artemyev}, \citenamefont
  {Vainchtein},\ and\ \citenamefont {Malova}}]{ZelenyiUFN2013}%
  \BibitemOpen
  \bibfield  {author} {\bibinfo {author} {\bibfnamefont {L.~M.}\ \bibnamefont
  {Zelenyi}}, \bibinfo {author} {\bibfnamefont {A.~I.}\ \bibnamefont
  {Neishtadt}}, \bibinfo {author} {\bibfnamefont {A.~V.}\ \bibnamefont
  {Artemyev}}, \bibinfo {author} {\bibfnamefont {D.~L.}\ \bibnamefont
  {Vainchtein}}, \ and\ \bibinfo {author} {\bibfnamefont {H.~V.}\ \bibnamefont
  {Malova}},\ }\href {\doibase 10.3367/UFNe.0183.201304b.0365} {\bibfield
  {journal} {\bibinfo  {journal} {Phys. Usp.}\ }\textbf {\bibinfo {volume}
  {56}},\ \bibinfo {pages} {347} (\bibinfo {year} {2013})}\BibitemShut
  {NoStop}%
\bibitem [{\citenamefont {Zelenyi}\ \emph {et~al.}(2016)\citenamefont
  {Zelenyi}, \citenamefont {Malova}, \citenamefont {Grigorenko},\ and\
  \citenamefont {Popov}}]{ZelenyiUFN2016}%
  \BibitemOpen
  \bibfield  {author} {\bibinfo {author} {\bibfnamefont {L.~M.}\ \bibnamefont
  {Zelenyi}}, \bibinfo {author} {\bibfnamefont {H.~V.}\ \bibnamefont {Malova}},
  \bibinfo {author} {\bibfnamefont {E.~E.}\ \bibnamefont {Grigorenko}}, \ and\
  \bibinfo {author} {\bibfnamefont {V.~Y.}\ \bibnamefont {Popov}},\ }\href
  {\doibase 10.3367/UFNe.2016.09.037923} {\bibfield  {journal} {\bibinfo
  {journal} {Phys. Usp.}\ }\textbf {\bibinfo {volume} {59}},\ \bibinfo {pages}
  {1057} (\bibinfo {year} {2016})}\BibitemShut {NoStop}%
\bibitem [{\citenamefont {Arons}(2012)}]{Arons2012}%
  \BibitemOpen
  \bibfield  {author} {\bibinfo {author} {\bibfnamefont {J.}~\bibnamefont
  {Arons}},\ }\href {\doibase https://doi.org/10.1007/s11214-012-9885-1}
  {\bibfield  {journal} {\bibinfo  {journal} {Space Science Reviews}\ }\textbf
  {\bibinfo {volume} {173}},\ \bibinfo {pages} {341} (\bibinfo {year}
  {2012})}\BibitemShut {NoStop}%
\bibitem [{\citenamefont {Hakobyan}\ \emph {et~al.}(2019)\citenamefont
  {Hakobyan}, \citenamefont {Philippov},\ and\ \citenamefont
  {Spitkovsky}}]{Phillipov}%
  \BibitemOpen
  \bibfield  {author} {\bibinfo {author} {\bibfnamefont {H.}~\bibnamefont
  {Hakobyan}}, \bibinfo {author} {\bibfnamefont {A.}~\bibnamefont {Philippov}},
  \ and\ \bibinfo {author} {\bibfnamefont {A.}~\bibnamefont {Spitkovsky}},\
  }\href {\doibase 10.3847/1538-4357/ab191b} {\bibfield  {journal} {\bibinfo
  {journal} {The Astrophysical Journal}\ }\textbf {\bibinfo {volume} {877}},\
  \bibinfo {pages} {53} (\bibinfo {year} {2019})}\BibitemShut {NoStop}%
\bibitem [{\citenamefont {Danson}\ \emph {et~al.}(2019)\citenamefont {Danson},
  \citenamefont {Haefner}, \citenamefont {Bromage}, \citenamefont {Butcher},
  \citenamefont {Chanteloup}, \citenamefont {Chowdhury}, \citenamefont
  {Galvanauskas}, \citenamefont {Gizzi}, \citenamefont {Hein}, \citenamefont
  {Hillier},\ and\ \citenamefont {et~al.}}]{MPWlasers}%
  \BibitemOpen
  \bibfield  {author} {\bibinfo {author} {\bibfnamefont {C.~N.}\ \bibnamefont
  {Danson}}, \bibinfo {author} {\bibfnamefont {C.}~\bibnamefont {Haefner}},
  \bibinfo {author} {\bibfnamefont {J.}~\bibnamefont {Bromage}}, \bibinfo
  {author} {\bibfnamefont {T.}~\bibnamefont {Butcher}}, \bibinfo {author}
  {\bibfnamefont {J.-C.~F.}\ \bibnamefont {Chanteloup}}, \bibinfo {author}
  {\bibfnamefont {E.~A.}\ \bibnamefont {Chowdhury}}, \bibinfo {author}
  {\bibfnamefont {A.}~\bibnamefont {Galvanauskas}}, \bibinfo {author}
  {\bibfnamefont {L.~A.}\ \bibnamefont {Gizzi}}, \bibinfo {author}
  {\bibfnamefont {J.}~\bibnamefont {Hein}}, \bibinfo {author} {\bibfnamefont
  {D.~I.}\ \bibnamefont {Hillier}}, \ and\ \bibinfo {author} {\bibnamefont
  {et~al.}},\ }\href {\doibase 10.1017/hpl.2019.36} {\bibfield  {journal}
  {\bibinfo  {journal} {High Power Laser Science and Engineering}\ }\textbf
  {\bibinfo {volume} {7}},\ \bibinfo {pages} {e54} (\bibinfo {year}
  {2019})}\BibitemShut {NoStop}%
\bibitem [{\citenamefont {Efimenko}\ \emph {et~al.}(2018)\citenamefont
  {Efimenko}, \citenamefont {Bashinov}, \citenamefont {Bastrakov},
  \citenamefont {Gonoskov}, \citenamefont {Muraviev}, \citenamefont {Meyerov},
  \citenamefont {Kim},\ and\ \citenamefont {Sergeev}}]{EfimenkoSR}%
  \BibitemOpen
  \bibfield  {author} {\bibinfo {author} {\bibfnamefont {E.~S.}\ \bibnamefont
  {Efimenko}}, \bibinfo {author} {\bibfnamefont {A.~V.}\ \bibnamefont
  {Bashinov}}, \bibinfo {author} {\bibfnamefont {S.~I.}\ \bibnamefont
  {Bastrakov}}, \bibinfo {author} {\bibfnamefont {A.~A.}\ \bibnamefont
  {Gonoskov}}, \bibinfo {author} {\bibfnamefont {A.~A.}\ \bibnamefont
  {Muraviev}}, \bibinfo {author} {\bibfnamefont {I.~B.}\ \bibnamefont
  {Meyerov}}, \bibinfo {author} {\bibfnamefont {A.~V.}\ \bibnamefont {Kim}}, \
  and\ \bibinfo {author} {\bibfnamefont {A.~M.}\ \bibnamefont {Sergeev}},\
  }\href {\doibase https://doi.org/10.1038/s41598-018-20745-y} {\bibfield
  {journal} {\bibinfo  {journal} {Scientific Reports}\ }\textbf {\bibinfo
  {volume} {8}},\ \bibinfo {pages} {2329} (\bibinfo {year} {2018})}\BibitemShut
  {NoStop}%
\bibitem [{\citenamefont {Muraviev}\ \emph {et~al.}(2015)\citenamefont
  {Muraviev}, \citenamefont {Bastrakov}, \citenamefont {Bashinov},
  \citenamefont {Gonoskov}, \citenamefont {Efimenko}, \citenamefont {Kim},
  \citenamefont {Meyerov},\ and\ \citenamefont {Sergeev}}]{MuravievJETPL}%
  \BibitemOpen
  \bibfield  {author} {\bibinfo {author} {\bibfnamefont {A.~A.}\ \bibnamefont
  {Muraviev}}, \bibinfo {author} {\bibfnamefont {S.~I.}\ \bibnamefont
  {Bastrakov}}, \bibinfo {author} {\bibfnamefont {A.~V.}\ \bibnamefont
  {Bashinov}}, \bibinfo {author} {\bibfnamefont {A.~A.}\ \bibnamefont
  {Gonoskov}}, \bibinfo {author} {\bibfnamefont {E.~S.}\ \bibnamefont
  {Efimenko}}, \bibinfo {author} {\bibfnamefont {A.~V.}\ \bibnamefont {Kim}},
  \bibinfo {author} {\bibfnamefont {I.~B.}\ \bibnamefont {Meyerov}}, \ and\
  \bibinfo {author} {\bibfnamefont {A.~M.}\ \bibnamefont {Sergeev}},\ }\href
  {\doibase https://doi.org/10.1134/S0021364015150060} {\bibfield  {journal}
  {\bibinfo  {journal} {Jetp Lett.}\ }\textbf {\bibinfo {volume} {102}},\
  \bibinfo {pages} {148} (\bibinfo {year} {2015})}\BibitemShut {NoStop}%
\bibitem [{\citenamefont {Bell}\ and\ \citenamefont {Kirk}(2008)}]{BellKirk}%
  \BibitemOpen
  \bibfield  {author} {\bibinfo {author} {\bibfnamefont {A.~R.}\ \bibnamefont
  {Bell}}\ and\ \bibinfo {author} {\bibfnamefont {J.~G.}\ \bibnamefont
  {Kirk}},\ }\href {\doibase 10.1103/PhysRevLett.101.200403} {\bibfield
  {journal} {\bibinfo  {journal} {Phys. Rev. Lett.}\ }\textbf {\bibinfo
  {volume} {101}},\ \bibinfo {pages} {200403} (\bibinfo {year}
  {2008})}\BibitemShut {NoStop}%
\bibitem [{\citenamefont {Gonoskov}\ \emph {et~al.}(2014)\citenamefont
  {Gonoskov}, \citenamefont {Bashinov}, \citenamefont {Gonoskov}, \citenamefont
  {Harvey}, \citenamefont {Ilderton}, \citenamefont {Kim}, \citenamefont
  {Marklund}, \citenamefont {Mourou},\ and\ \citenamefont {Sergeev}}]{ART}%
  \BibitemOpen
  \bibfield  {author} {\bibinfo {author} {\bibfnamefont {A.}~\bibnamefont
  {Gonoskov}}, \bibinfo {author} {\bibfnamefont {A.}~\bibnamefont {Bashinov}},
  \bibinfo {author} {\bibfnamefont {I.}~\bibnamefont {Gonoskov}}, \bibinfo
  {author} {\bibfnamefont {C.}~\bibnamefont {Harvey}}, \bibinfo {author}
  {\bibfnamefont {A.}~\bibnamefont {Ilderton}}, \bibinfo {author}
  {\bibfnamefont {A.}~\bibnamefont {Kim}}, \bibinfo {author} {\bibfnamefont
  {M.}~\bibnamefont {Marklund}}, \bibinfo {author} {\bibfnamefont
  {G.}~\bibnamefont {Mourou}}, \ and\ \bibinfo {author} {\bibfnamefont
  {A.}~\bibnamefont {Sergeev}},\ }\href {\doibase
  10.1103/PhysRevLett.113.014801} {\bibfield  {journal} {\bibinfo  {journal}
  {Phys. Rev. Lett.}\ }\textbf {\bibinfo {volume} {113}},\ \bibinfo {pages}
  {014801} (\bibinfo {year} {2014})}\BibitemShut {NoStop}%
\bibitem [{\citenamefont {Bashinov}\ \emph {et~al.}(2014)\citenamefont
  {Bashinov}, \citenamefont {Gonoskov}, \citenamefont {Kim}, \citenamefont
  {Mourou},\ and\ \citenamefont {Sergeev}}]{EPJ}%
  \BibitemOpen
  \bibfield  {author} {\bibinfo {author} {\bibfnamefont {A.}~\bibnamefont
  {Bashinov}}, \bibinfo {author} {\bibfnamefont {A.}~\bibnamefont {Gonoskov}},
  \bibinfo {author} {\bibfnamefont {A.}~\bibnamefont {Kim}}, \bibinfo {author}
  {\bibfnamefont {G.}~\bibnamefont {Mourou}}, \ and\ \bibinfo {author}
  {\bibfnamefont {A.}~\bibnamefont {Sergeev}},\ }\href {\doibase
  https://doi.org/10.1140/epjst/e2014-02161-7} {\bibfield  {journal} {\bibinfo
  {journal} {Eur. Phys. J. Spec. Top.}\ }\textbf {\bibinfo {volume} {223}},\
  \bibinfo {pages} {1105} (\bibinfo {year} {2014})}\BibitemShut {NoStop}%
\bibitem [{\citenamefont {Ji}\ \emph {et~al.}(2014)\citenamefont {Ji},
  \citenamefont {Pukhov}, \citenamefont {Kostyukov}, \citenamefont {Shen},\
  and\ \citenamefont {Akli}}]{PukhovPRL}%
  \BibitemOpen
  \bibfield  {author} {\bibinfo {author} {\bibfnamefont {L.~L.}\ \bibnamefont
  {Ji}}, \bibinfo {author} {\bibfnamefont {A.}~\bibnamefont {Pukhov}}, \bibinfo
  {author} {\bibfnamefont {I.~Y.}\ \bibnamefont {Kostyukov}}, \bibinfo {author}
  {\bibfnamefont {B.~F.}\ \bibnamefont {Shen}}, \ and\ \bibinfo {author}
  {\bibfnamefont {K.}~\bibnamefont {Akli}},\ }\href {\doibase
  10.1103/PhysRevLett.112.145003} {\bibfield  {journal} {\bibinfo  {journal}
  {Phys. Rev. Lett.}\ }\textbf {\bibinfo {volume} {112}},\ \bibinfo {pages}
  {145003} (\bibinfo {year} {2014})}\BibitemShut {NoStop}%
\bibitem [{\citenamefont {Neitz}\ and\ \citenamefont
  {Di~Piazza}(2013)}]{NeitzPRL}%
  \BibitemOpen
  \bibfield  {author} {\bibinfo {author} {\bibfnamefont {N.}~\bibnamefont
  {Neitz}}\ and\ \bibinfo {author} {\bibfnamefont {A.}~\bibnamefont
  {Di~Piazza}},\ }\href {\doibase 10.1103/PhysRevLett.111.054802} {\bibfield
  {journal} {\bibinfo  {journal} {Phys. Rev. Lett.}\ }\textbf {\bibinfo
  {volume} {111}},\ \bibinfo {pages} {054802} (\bibinfo {year}
  {2013})}\BibitemShut {NoStop}%
\bibitem [{\citenamefont {Zel'dovich}(1975)}]{Zeld}%
  \BibitemOpen
  \bibfield  {author} {\bibinfo {author} {\bibfnamefont {Y.~B.}\ \bibnamefont
  {Zel'dovich}},\ }\href@noop {} {\bibfield  {journal} {\bibinfo  {journal}
  {Sov. Phys. Usp.}\ }\textbf {\bibinfo {volume} {18}},\ \bibinfo {pages} {79}
  (\bibinfo {year} {1975})}\BibitemShut {NoStop}%
\bibitem [{\citenamefont {Bulanov}\ \emph {et~al.}(2004)\citenamefont
  {Bulanov}, \citenamefont {Esirkepov}, \citenamefont {Koga},\ and\
  \citenamefont {Tajima}}]{BulanovPPR}%
  \BibitemOpen
  \bibfield  {author} {\bibinfo {author} {\bibfnamefont {S.~V.}\ \bibnamefont
  {Bulanov}}, \bibinfo {author} {\bibfnamefont {T.~Z.}\ \bibnamefont
  {Esirkepov}}, \bibinfo {author} {\bibfnamefont {J.}~\bibnamefont {Koga}}, \
  and\ \bibinfo {author} {\bibfnamefont {T.}~\bibnamefont {Tajima}},\ }\href
  {\doibase https://doi.org/10.1134/1.1687021} {\bibfield  {journal} {\bibinfo
  {journal} {Plasma Phys. Rep.}\ }\textbf {\bibinfo {volume} {30}},\ \bibinfo
  {pages} {196} (\bibinfo {year} {2004})}\BibitemShut {NoStop}%
\bibitem [{\citenamefont {Kirk}(2016)}]{Kirk_2016}%
  \BibitemOpen
  \bibfield  {author} {\bibinfo {author} {\bibfnamefont {J.~G.}\ \bibnamefont
  {Kirk}},\ }\href {\doibase 10.1088/0741-3335/58/8/085005} {\bibfield
  {journal} {\bibinfo  {journal} {Plasma Physics and Controlled Fusion}\
  }\textbf {\bibinfo {volume} {58}},\ \bibinfo {pages} {085005} (\bibinfo
  {year} {2016})}\BibitemShut {NoStop}%
\bibitem [{\citenamefont {Gelfer}\ \emph {et~al.}(2018)\citenamefont {Gelfer},
  \citenamefont {Elkina},\ and\ \citenamefont {Fedotov}}]{GelferSR}%
  \BibitemOpen
  \bibfield  {author} {\bibinfo {author} {\bibfnamefont {E.}~\bibnamefont
  {Gelfer}}, \bibinfo {author} {\bibfnamefont {N.}~\bibnamefont {Elkina}}, \
  and\ \bibinfo {author} {\bibfnamefont {A.}~\bibnamefont {Fedotov}},\ }\href
  {\doibase https://doi.org/10.1038/s41598-018-24930-x} {\bibfield  {journal}
  {\bibinfo  {journal} {Sci Rep}\ }\textbf {\bibinfo {volume} {8}},\ \bibinfo
  {pages} {6478} (\bibinfo {year} {2018})}\BibitemShut {NoStop}%
\bibitem [{\citenamefont {Jaroschek}\ and\ \citenamefont
  {Hoshino}(2009)}]{Jaroschek2009}%
  \BibitemOpen
  \bibfield  {author} {\bibinfo {author} {\bibfnamefont {C.~H.}\ \bibnamefont
  {Jaroschek}}\ and\ \bibinfo {author} {\bibfnamefont {M.}~\bibnamefont
  {Hoshino}},\ }\href {\doibase 10.1103/PhysRevLett.103.075002} {\bibfield
  {journal} {\bibinfo  {journal} {Phys. Rev. Lett.}\ }\textbf {\bibinfo
  {volume} {103}},\ \bibinfo {pages} {075002} (\bibinfo {year}
  {2009})}\BibitemShut {NoStop}%
\bibitem [{\citenamefont {Runov}\ \emph {et~al.}(2003)\citenamefont {Runov},
  \citenamefont {Nakamura}, \citenamefont {Baumjohann}, \citenamefont
  {Treumann}, \citenamefont {Zhang}, \citenamefont {Volwerk}, \citenamefont
  {Vörös}, \citenamefont {Balogh}, \citenamefont {Glaßmeier}, \citenamefont
  {Klecker}, \citenamefont {Rème},\ and\ \citenamefont {Kistler}}]{Runov2003}%
  \BibitemOpen
  \bibfield  {author} {\bibinfo {author} {\bibfnamefont {A.}~\bibnamefont
  {Runov}}, \bibinfo {author} {\bibfnamefont {R.}~\bibnamefont {Nakamura}},
  \bibinfo {author} {\bibfnamefont {W.}~\bibnamefont {Baumjohann}}, \bibinfo
  {author} {\bibfnamefont {R.~A.}\ \bibnamefont {Treumann}}, \bibinfo {author}
  {\bibfnamefont {T.~L.}\ \bibnamefont {Zhang}}, \bibinfo {author}
  {\bibfnamefont {M.}~\bibnamefont {Volwerk}}, \bibinfo {author} {\bibfnamefont
  {Z.}~\bibnamefont {Vörös}}, \bibinfo {author} {\bibfnamefont
  {A.}~\bibnamefont {Balogh}}, \bibinfo {author} {\bibfnamefont {K.-H.}\
  \bibnamefont {Glaßmeier}}, \bibinfo {author} {\bibfnamefont
  {B.}~\bibnamefont {Klecker}}, \bibinfo {author} {\bibfnamefont
  {H.}~\bibnamefont {Rème}}, \ and\ \bibinfo {author} {\bibfnamefont
  {L.}~\bibnamefont {Kistler}},\ }\href {\doibase
  https://doi.org/10.1029/2002GL016730} {\bibfield  {journal} {\bibinfo
  {journal} {Geophysical Research Letters}\ }\textbf {\bibinfo {volume} {30}}
  (\bibinfo {year} {2003}),\ https://doi.org/10.1029/2002GL016730}\BibitemShut
  {NoStop}%
\bibitem [{\citenamefont {Landau}\ and\ \citenamefont {Lifshitz}(1975)}]{LL}%
  \BibitemOpen
  \bibfield  {author} {\bibinfo {author} {\bibfnamefont {L.~D.}\ \bibnamefont
  {Landau}}\ and\ \bibinfo {author} {\bibfnamefont {E.~M.}\ \bibnamefont
  {Lifshitz}},\ }\href@noop {} {\emph {\bibinfo {title} {The Classical Theory
  of Fields}}}\ (\bibinfo  {publisher} {Elsevier, Oxford},\ \bibinfo {year}
  {1975})\BibitemShut {NoStop}%
\bibitem [{\citenamefont {Berestetskii}\ \emph {et~al.}(1982)\citenamefont
  {Berestetskii}, \citenamefont {Lifshits},\ and\ \citenamefont
  {Pitaevskii}}]{BLP}%
  \BibitemOpen
  \bibfield  {author} {\bibinfo {author} {\bibfnamefont {V.~B.}\ \bibnamefont
  {Berestetskii}}, \bibinfo {author} {\bibfnamefont {E.~M.}\ \bibnamefont
  {Lifshits}}, \ and\ \bibinfo {author} {\bibfnamefont {L.~P.}\ \bibnamefont
  {Pitaevskii}},\ }\href@noop {} {\emph {\bibinfo {title} {Quantum
  Electrodynamics}}}\ (\bibinfo  {publisher} {Pergamon Press, New York},\
  \bibinfo {year} {1982})\BibitemShut {NoStop}%
\bibitem [{\citenamefont {Bayer}\ \emph {et~al.}(1973)\citenamefont {Bayer},
  \citenamefont {Katkov},\ and\ \citenamefont {Fadin}}]{BKF}%
  \BibitemOpen
  \bibfield  {author} {\bibinfo {author} {\bibfnamefont {V.~N.}\ \bibnamefont
  {Bayer}}, \bibinfo {author} {\bibfnamefont {V.~M.}\ \bibnamefont {Katkov}}, \
  and\ \bibinfo {author} {\bibfnamefont {V.~S.}\ \bibnamefont {Fadin}},\
  }\href@noop {} {\emph {\bibinfo {title} {Radiation of the Relativistic
  Electrons}}}\ (\bibinfo  {publisher} {Atomizdat, Moscow},\ \bibinfo {year}
  {1973})\BibitemShut {NoStop}%
\bibitem [{\citenamefont {Nikishov}\ and\ \citenamefont
  {Ritus}(1964{\natexlab{a}})}]{NikishovRitus}%
  \BibitemOpen
  \bibfield  {author} {\bibinfo {author} {\bibfnamefont {A.~I.}\ \bibnamefont
  {Nikishov}}\ and\ \bibinfo {author} {\bibfnamefont {V.~I.}\ \bibnamefont
  {Ritus}},\ }\href@noop {} {\bibfield  {journal} {\bibinfo  {journal} {Sov.
  Phys. JETP}\ }\textbf {\bibinfo {volume} {19}},\ \bibinfo {pages} {529}
  (\bibinfo {year} {1964}{\natexlab{a}})}\BibitemShut {NoStop}%
\bibitem [{\citenamefont {Nikishov}\ and\ \citenamefont
  {Ritus}(1964{\natexlab{b}})}]{NikishovRitusII}%
  \BibitemOpen
  \bibfield  {author} {\bibinfo {author} {\bibfnamefont {A.~I.}\ \bibnamefont
  {Nikishov}}\ and\ \bibinfo {author} {\bibfnamefont {V.~I.}\ \bibnamefont
  {Ritus}},\ }\href@noop {} {\bibfield  {journal} {\bibinfo  {journal} {Sov.
  Phys. JETP}\ }\textbf {\bibinfo {volume} {19}},\ \bibinfo {pages} {1191}
  (\bibinfo {year} {1964}{\natexlab{b}})}\BibitemShut {NoStop}%
\bibitem [{\citenamefont {Poder}\ \emph {et~al.}(2018)\citenamefont {Poder},
  \citenamefont {Tamburini}, \citenamefont {Sarri}, \citenamefont {Di~Piazza},
  \citenamefont {Kuschel}, \citenamefont {Baird}, \citenamefont {Behm},
  \citenamefont {Bohlen}, \citenamefont {Cole}, \citenamefont {Corvan},
  \citenamefont {Duff}, \citenamefont {Gerstmayr}, \citenamefont {Keitel},
  \citenamefont {Krushelnick}, \citenamefont {Mangles}, \citenamefont
  {McKenna}, \citenamefont {Murphy}, \citenamefont {Najmudin}, \citenamefont
  {Ridgers}, \citenamefont {Samarin}, \citenamefont {Symes}, \citenamefont
  {Thomas}, \citenamefont {Warwick},\ and\ \citenamefont {Zepf}}]{PoderPRX}%
  \BibitemOpen
  \bibfield  {author} {\bibinfo {author} {\bibfnamefont {K.}~\bibnamefont
  {Poder}}, \bibinfo {author} {\bibfnamefont {M.}~\bibnamefont {Tamburini}},
  \bibinfo {author} {\bibfnamefont {G.}~\bibnamefont {Sarri}}, \bibinfo
  {author} {\bibfnamefont {A.}~\bibnamefont {Di~Piazza}}, \bibinfo {author}
  {\bibfnamefont {S.}~\bibnamefont {Kuschel}}, \bibinfo {author} {\bibfnamefont
  {C.~D.}\ \bibnamefont {Baird}}, \bibinfo {author} {\bibfnamefont
  {K.}~\bibnamefont {Behm}}, \bibinfo {author} {\bibfnamefont {S.}~\bibnamefont
  {Bohlen}}, \bibinfo {author} {\bibfnamefont {J.~M.}\ \bibnamefont {Cole}},
  \bibinfo {author} {\bibfnamefont {D.~J.}\ \bibnamefont {Corvan}}, \bibinfo
  {author} {\bibfnamefont {M.}~\bibnamefont {Duff}}, \bibinfo {author}
  {\bibfnamefont {E.}~\bibnamefont {Gerstmayr}}, \bibinfo {author}
  {\bibfnamefont {C.~H.}\ \bibnamefont {Keitel}}, \bibinfo {author}
  {\bibfnamefont {K.}~\bibnamefont {Krushelnick}}, \bibinfo {author}
  {\bibfnamefont {S.~P.~D.}\ \bibnamefont {Mangles}}, \bibinfo {author}
  {\bibfnamefont {P.}~\bibnamefont {McKenna}}, \bibinfo {author} {\bibfnamefont
  {C.~D.}\ \bibnamefont {Murphy}}, \bibinfo {author} {\bibfnamefont
  {Z.}~\bibnamefont {Najmudin}}, \bibinfo {author} {\bibfnamefont {C.~P.}\
  \bibnamefont {Ridgers}}, \bibinfo {author} {\bibfnamefont {G.~M.}\
  \bibnamefont {Samarin}}, \bibinfo {author} {\bibfnamefont {D.~R.}\
  \bibnamefont {Symes}}, \bibinfo {author} {\bibfnamefont {A.~G.~R.}\
  \bibnamefont {Thomas}}, \bibinfo {author} {\bibfnamefont {J.}~\bibnamefont
  {Warwick}}, \ and\ \bibinfo {author} {\bibfnamefont {M.}~\bibnamefont
  {Zepf}},\ }\href {\doibase 10.1103/PhysRevX.8.031004} {\bibfield  {journal}
  {\bibinfo  {journal} {Phys. Rev. X}\ }\textbf {\bibinfo {volume} {8}},\
  \bibinfo {pages} {031004} (\bibinfo {year} {2018})}\BibitemShut {NoStop}%
\bibitem [{\citenamefont {Cole}\ \emph {et~al.}(2018)\citenamefont {Cole},
  \citenamefont {Behm}, \citenamefont {Gerstmayr}, \citenamefont {Blackburn},
  \citenamefont {Wood}, \citenamefont {Baird}, \citenamefont {Duff},
  \citenamefont {Harvey}, \citenamefont {Ilderton}, \citenamefont {Joglekar},
  \citenamefont {Krushelnick}, \citenamefont {Kuschel}, \citenamefont
  {Marklund}, \citenamefont {McKenna}, \citenamefont {Murphy}, \citenamefont
  {Poder}, \citenamefont {Ridgers}, \citenamefont {Samarin}, \citenamefont
  {Sarri}, \citenamefont {Symes}, \citenamefont {Thomas}, \citenamefont
  {Warwick}, \citenamefont {Zepf}, \citenamefont {Najmudin},\ and\
  \citenamefont {Mangles}}]{ColePRX}%
  \BibitemOpen
  \bibfield  {author} {\bibinfo {author} {\bibfnamefont {J.~M.}\ \bibnamefont
  {Cole}}, \bibinfo {author} {\bibfnamefont {K.~T.}\ \bibnamefont {Behm}},
  \bibinfo {author} {\bibfnamefont {E.}~\bibnamefont {Gerstmayr}}, \bibinfo
  {author} {\bibfnamefont {T.~G.}\ \bibnamefont {Blackburn}}, \bibinfo {author}
  {\bibfnamefont {J.~C.}\ \bibnamefont {Wood}}, \bibinfo {author}
  {\bibfnamefont {C.~D.}\ \bibnamefont {Baird}}, \bibinfo {author}
  {\bibfnamefont {M.~J.}\ \bibnamefont {Duff}}, \bibinfo {author}
  {\bibfnamefont {C.}~\bibnamefont {Harvey}}, \bibinfo {author} {\bibfnamefont
  {A.}~\bibnamefont {Ilderton}}, \bibinfo {author} {\bibfnamefont {A.~S.}\
  \bibnamefont {Joglekar}}, \bibinfo {author} {\bibfnamefont {K.}~\bibnamefont
  {Krushelnick}}, \bibinfo {author} {\bibfnamefont {S.}~\bibnamefont
  {Kuschel}}, \bibinfo {author} {\bibfnamefont {M.}~\bibnamefont {Marklund}},
  \bibinfo {author} {\bibfnamefont {P.}~\bibnamefont {McKenna}}, \bibinfo
  {author} {\bibfnamefont {C.~D.}\ \bibnamefont {Murphy}}, \bibinfo {author}
  {\bibfnamefont {K.}~\bibnamefont {Poder}}, \bibinfo {author} {\bibfnamefont
  {C.~P.}\ \bibnamefont {Ridgers}}, \bibinfo {author} {\bibfnamefont {G.~M.}\
  \bibnamefont {Samarin}}, \bibinfo {author} {\bibfnamefont {G.}~\bibnamefont
  {Sarri}}, \bibinfo {author} {\bibfnamefont {D.~R.}\ \bibnamefont {Symes}},
  \bibinfo {author} {\bibfnamefont {A.~G.~R.}\ \bibnamefont {Thomas}}, \bibinfo
  {author} {\bibfnamefont {J.}~\bibnamefont {Warwick}}, \bibinfo {author}
  {\bibfnamefont {M.}~\bibnamefont {Zepf}}, \bibinfo {author} {\bibfnamefont
  {Z.}~\bibnamefont {Najmudin}}, \ and\ \bibinfo {author} {\bibfnamefont
  {S.~P.~D.}\ \bibnamefont {Mangles}},\ }\href {\doibase
  10.1103/PhysRevX.8.011020} {\bibfield  {journal} {\bibinfo  {journal} {Phys.
  Rev. X}\ }\textbf {\bibinfo {volume} {8}},\ \bibinfo {pages} {011020}
  (\bibinfo {year} {2018})}\BibitemShut {NoStop}%
\bibitem [{\citenamefont {Wistisen}\ \emph {et~al.}(2018)\citenamefont
  {Wistisen}, \citenamefont {Piazza}, \citenamefont {Knudsen},\ and\
  \citenamefont {Uggerhøj}}]{Wistisen}%
  \BibitemOpen
  \bibfield  {author} {\bibinfo {author} {\bibfnamefont {T.~N.}\ \bibnamefont
  {Wistisen}}, \bibinfo {author} {\bibfnamefont {A.~D.}\ \bibnamefont
  {Piazza}}, \bibinfo {author} {\bibfnamefont {H.~V.}\ \bibnamefont {Knudsen}},
  \ and\ \bibinfo {author} {\bibfnamefont {U.~I.}\ \bibnamefont {Uggerhøj}},\
  }\href {\doibase https://doi.org/10.1038/s41467-018-03165-4} {\bibfield
  {journal} {\bibinfo  {journal} {Nat Commun}\ }\textbf {\bibinfo {volume}
  {9}},\ \bibinfo {pages} {795} (\bibinfo {year} {2018})}\BibitemShut {NoStop}%
\bibitem [{\citenamefont {Bashinov}\ \emph {et~al.}(2015)\citenamefont
  {Bashinov}, \citenamefont {Kim},\ and\ \citenamefont
  {Sergeev}}]{BashinovPRE}%
  \BibitemOpen
  \bibfield  {author} {\bibinfo {author} {\bibfnamefont {A.~V.}\ \bibnamefont
  {Bashinov}}, \bibinfo {author} {\bibfnamefont {A.~V.}\ \bibnamefont {Kim}}, \
  and\ \bibinfo {author} {\bibfnamefont {A.~M.}\ \bibnamefont {Sergeev}},\
  }\href {\doibase 10.1103/PhysRevE.92.043105} {\bibfield  {journal} {\bibinfo
  {journal} {Phys. Rev. E}\ }\textbf {\bibinfo {volume} {92}},\ \bibinfo
  {pages} {043105} (\bibinfo {year} {2015})}\BibitemShut {NoStop}%
\bibitem [{\citenamefont {Bashinov}\ \emph {et~al.}(2017)\citenamefont
  {Bashinov}, \citenamefont {Kumar},\ and\ \citenamefont {Kim}}]{BashinovPRA}%
  \BibitemOpen
  \bibfield  {author} {\bibinfo {author} {\bibfnamefont {A.~V.}\ \bibnamefont
  {Bashinov}}, \bibinfo {author} {\bibfnamefont {P.}~\bibnamefont {Kumar}}, \
  and\ \bibinfo {author} {\bibfnamefont {A.~V.}\ \bibnamefont {Kim}},\ }\href
  {\doibase 10.1103/PhysRevA.95.042127} {\bibfield  {journal} {\bibinfo
  {journal} {Phys. Rev. A}\ }\textbf {\bibinfo {volume} {95}},\ \bibinfo
  {pages} {042127} (\bibinfo {year} {2017})}\BibitemShut {NoStop}%
\bibitem [{Note1()}]{Note1}%
  \BibitemOpen
  \bibinfo {note} {This particular expression can be derived with more ease by
  assuming $\protect \qopname \relax o{sin}{\varphi }=\varphi $ in (\ref
  {systemradsubstphi}) and then proceeding with (\ref {eq:x})}\BibitemShut
  {NoStop}%
\bibitem [{Note2()}]{Note2}%
  \BibitemOpen
  \bibinfo {note} {The map for the parameter $\protect \frac {p}{mc}$ happens
  to be very similar, so we chose to omit it.}\BibitemShut {Stop}%
\bibitem [{\citenamefont {Esirkepov}\ \emph {et~al.}(2015)\citenamefont
  {Esirkepov}, \citenamefont {Bulanov}, \citenamefont {Koga}, \citenamefont
  {Kando}, \citenamefont {Kondo}, \citenamefont {Rosanov}, \citenamefont
  {Korn},\ and\ \citenamefont {Bulanov}}]{EsirkepovPLA}%
  \BibitemOpen
  \bibfield  {author} {\bibinfo {author} {\bibfnamefont {T.~Z.}\ \bibnamefont
  {Esirkepov}}, \bibinfo {author} {\bibfnamefont {S.~S.}\ \bibnamefont
  {Bulanov}}, \bibinfo {author} {\bibfnamefont {J.~K.}\ \bibnamefont {Koga}},
  \bibinfo {author} {\bibfnamefont {M.}~\bibnamefont {Kando}}, \bibinfo
  {author} {\bibfnamefont {K.}~\bibnamefont {Kondo}}, \bibinfo {author}
  {\bibfnamefont {N.~N.}\ \bibnamefont {Rosanov}}, \bibinfo {author}
  {\bibfnamefont {G.}~\bibnamefont {Korn}}, \ and\ \bibinfo {author}
  {\bibfnamefont {S.~V.}\ \bibnamefont {Bulanov}},\ }\href {\doibase
  https://doi.org/10.1016/j.physleta.2015.06.017} {\bibfield  {journal}
  {\bibinfo  {journal} {Physics Letters A}\ }\textbf {\bibinfo {volume}
  {379}},\ \bibinfo {pages} {2044} (\bibinfo {year} {2015})}\BibitemShut
  {NoStop}%
\bibitem [{\citenamefont {Erber}(1966)}]{erber.rmp.1966}%
  \BibitemOpen
  \bibfield  {author} {\bibinfo {author} {\bibfnamefont {T.}~\bibnamefont
  {Erber}},\ }\href {\doibase 10.1103/RevModPhys.38.626} {\bibfield  {journal}
  {\bibinfo  {journal} {Reviews of Modern Physics}\ }\textbf {\bibinfo {volume}
  {38}},\ \bibinfo {pages} {626} (\bibinfo {year} {1966})}\BibitemShut
  {NoStop}%
\end{thebibliography}%

\end{document}